\documentclass[%
 aip,
 amsmath,amssymb,
 reprint,%
]{revtex4-1}

\usepackage{graphicx}
\usepackage{dcolumn}
\usepackage{bm}
\usepackage[utf8]{inputenc}
\usepackage[T1]{fontenc}
\usepackage{mathptmx}
\usepackage{color,soul}
\usepackage{amssymb}
\usepackage{amsmath}
\usepackage{longtable,tabularx}
\usepackage{xcolor}
\usepackage{subfloat}
\usepackage{multirow}
\usepackage{rotating}
\usepackage{booktabs}
\usepackage{relsize}
\usepackage{nomencl}
\usepackage{etoolbox}
\usepackage{float}
\usepackage{url}
\usepackage{hyperref}
\usepackage[ruled]{algorithm2e}
\usepackage{nomencl}
\usepackage{xcolor}
\usepackage[normalem]{ulem}

\newcommand{\pder}[2][]{\frac{\partial#1}{\partial#2}}
\newcommand{\der}[2][]{\frac{d#1}{d#2}}

\makenomenclature
\begin{document}

\preprint{AIP/123-QED}

\title{Predicting Waves in Fluids with Deep Neural Network}
\affiliation{Department of Mechanical Engineering, University of British Columbia, V6T 1Z4, Canada}
\author{Indu Kant Deo}
\email{indukant@mail.ubc.ca}

\author{Rajeev Jaiman}
\email{rjaiman@mech.ubc.ca}

\date{\today}

\begin{abstract}
In this paper, we present a deep learning technique for data-driven predictions of wave propagation in a fluid medium. 
The technique relies on an attention-based convolutional recurrent autoencoder network (AB-CRAN).
To construct a low-dimensional representation of wave propagation data, we employ a denoising-based convolutional autoencoder.
The AB-CRAN architecture with attention-based long short-term memory cells forms our deep neural network model for the time marching of the low-dimensional features. 
We assess the proposed AB-CRAN framework  against the standard recurrent neural network for the low-dimensional learning of wave propagation. 
To demonstrate the effectiveness of the AB-CRAN model, we consider three benchmark problems, namely one-dimensional linear convection, the nonlinear viscous Burgers equation, and the two-dimensional Saint-Venant shallow water system.
Using the spatial-temporal datasets from the benchmark problems, our novel AB-CRAN architecture accurately captures the wave amplitude and preserves the wave characteristics of the solution for long time horizons. 
The attention-based sequence-to-sequence network increases the time-horizon of prediction compared to the standard recurrent neural network with long short-term memory cells. The denoising autoencoder further reduces the mean squared error of prediction and improves the generalization capability in the parameter space.

\keywords{Wave propagation; Data-driven prediction; Convolutional recurrent autoencoder network; Denoising autoencoders;  Attention mechanism; Sequence-to-sequence modeling;} 
\end{abstract}

\maketitle

\section{Introduction}
A wide range of physical phenomena involving wave motion and convective transport processes can be modeled using hyperbolic systems of partial differential equations \cite{leveque2002finite}.
Efficient and accurate solutions to these high-dimensional partial differential equations are critical in a variety of scientific and engineering disciplines ranging from shock waves in fluids, shallow water waves to underwater noise propagation \cite{lighthill2001waves}. 
However, solving these high-dimensional partial differential equations using state-of-the-art numerical discretization techniques can be computationally expensive, making them unsuitable for multi-query parametric analysis, design optimization, and control-related tasks.
For example, there is a practical need for efficient large-scale prediction, decision making, and effective mitigation strategies for underwater radiated noise caused by various human activities such as ship traffic, sonar and seismic air gun blasts \cite{erbe2019effects}. 
Anthropogenic noise is shown to adversely impact marine animals~\cite{duarte2021soundscape}, which is a major concern for both environmentalists and the offshore/shipbuilding industry. 
Underwater acoustic noise prediction from widely varying sources forms a parameter driven problem that necessitates an efficient data-driven modeling capability to scan a large parameter space. 
Specifically, this work is motivated by the development of a framework for parametric data-driven predictions of wave propagation and convection-dominated physical phenomena.

Reduced-order modeling has emerged as a valuable tool in the design of large-scale systems while dealing with multi-query analysis and optimization.
Reduced-order models attempt to replace the full-order model with a lower-dimensional representation capable of expressing the physical properties of the problem described by the full-order model. 
{This is achieved by finding low-rank structures in the full-order data that describe the underlying spatial-temporal dynamics.
One type of reduced-order modeling technique is based on the idea that a reduced-order approximation can be described as a linear combination of basis functions constructed from a series of full-order model solutions known as snapshots \cite{schilders2008model}. 
To construct an optimal low-rank approximation of the solution manifold, one of these methodologies is proper orthogonal decomposition (POD), which generates a linear reduced-order model by decomposing a snapshot matrix into singular values \cite{quarteroni2015reduced}.
The use of the proper orthogonal decomposition to find a reduced subspace and the Galerkin projection \cite{DBLP:journals/corr/CarlbergBA15} for evolving dynamics in this reduced space is a typical approach for building a reduced-order model for a dynamical system with known governing equations.} 
However, POD-based reduced-order model is inefficient for practical problems where the worst-case error from the best-approximated linear subspace decays slowly with increasing subspace dimension. This is often the case in the low-dimensional representation when dealing with wave propagation and convection-dominated problems via POD-based reduced-order model.

{It is well established in the literature that there are fundamental challenges to reducing the dimensionality of convection-dominated problems governed by hyperbolic PDEs using linear subspace and traditional projection-based reduced-order models \cite{unger2019kolmogorov,greif2019decay,cagniart2019model}.
To address this difficulty,  several nonlinear manifold learning methods are being developed for improved mode extraction and low-rank representations such as iso-map \cite{tenenbaum1998mapping}, kernel principal component analysis \cite{mika1998kernel, salvador2021non}, and diffeomorphic dimensionality reduction \cite{seeger2009advances,mojgani2020physics}.} 
These methods attempt to reduce the dimensionality of the convection-dominated problem (i.e., wave propagation) and construct a nonlinear reduced-order model. 
Although recently nonlinear manifold learning approaches have been demonstrated with some success in reducing the dimensionality of convection dominated data \cite{mojgani2020physics,salvador2021non}, they often lack simplicity and generality in mapping the data from the low-dimensional space to the high-dimensional physical space. 
As a result, many of these methods are usually not the preferred choice in the engineering community when it comes to reduced-order modeling of hyperbolic PDE systems. Using machine learning and deep neural networks, the goal of this research is to develop a data-driven reduced-order model for wave propagation and convection-dominated problems.

In recent years, machine learning has witnessed a resurgence with groundbreaking successes in a wide range of technological and engineering applications~\cite{lecun2015deep}. 
Reduced-order modeling is one type of application, in which deep learning models can be used to approximate a physical system in low-dimension and make an inference \cite{gupta2022three}. 
For example, using deep learning algorithms, one can construct data-driven reduced-order models of fluid flow \cite{miyanawala2017efficient,miyanawala2018novel} and nonlinear dynamical systems e.g., fluid-structure interaction \cite{miyanawala2018low}.
By extracting fluid features such as vortices and learning the evolution of these vortices from spatial-temporal data, deep learning models can learn complex relationships and patterns for fast and efficient inference. 
Deep neural networks are proven to be an extremely effective method for a low-dimensional representation of a wide range of physical systems such as Navier-Stokes equations~\cite{yang2016data}, turbulence modelling~\cite{geneva2019quantifying}, and among others. 
However, it is widely acknowledged that the training of deep learning models may require a substantial amount of data and computational time. There is also a growing interest to incorporate prior knowledge into these models to improve interpretability. Some prior domain knowledge and inductive biases \cite{goyal2020inductive} can provide interpretability and help reduce the amount of data required in these deep learning models. This intent has resulted in the development of physics-guided neural networks, which incorporate domain-specific physical information into deep learning architectures~\cite{willard2021integrating}.

With the advent of deep learning, low-dimensional representation learning via neutral networks  is being explored to overcome the limitations of nonlinear manifold learning in mapping low-dimensional space to high-dimensional physical space \cite{hinton2006reducing}.
For example, autoencoders allow compressing the input data to a low-dimensional space via encoder network and back to the high-dimensional physical space using decoder network \cite{lee2020model, gonzalez2018deep,bukka2020deep,sorteberg2018approximating,fresca2021comprehensive, maulik2021reduced, XU2020113379}.
Using a series of convolutional and nonlinear mapping operations, autoencoders can provide an efficient construction of a compressed low-dimensional space. 
{A comprehensive overview of autoencoders and their various types can be found in \cite{goodfellow2016deep,dong2018review}.}
%
The autoencoders can be interpreted as a flexible and nonlinear generalization of proper orthogonal decomposition \cite{plaut2018principal,bukka2021assessment}.
Using the proper autoencoder architecture, one can circumvent some of the drawbacks of the POD-based reduced-order models for constructing the reduced-order space of the snapshot matrix. Similar to POD, the subspace projection in an autoencoder is achieved using an unsupervised training of the snapshot matrix. As shown in \cite{plaut2018principal}, a linear autoencoder recovers the reduced subspace spanned by the POD after the convergence. Unlike the POD, the modes for the autoencoders are not orthogonal and the orthogonality can be imparted on the weights during the training by performing a singular value decomposition \cite{plaut2018principal}.

%
Deep neural network architectures, such as convolutional recurrent autoencoders, are efficient and useful for constructing low-dimensional learning models for data-driven modelling of nonlinear PDEs \cite{gonzalez2018deep,bukka2021assessment,gupta2022hybrid}.
A neural network based on the convolutional recurrent autoencoder provides a fully data-driven framework in which deep learning algorithms are used to learn both the low-dimensional representation of the state and its time evolution.
Convolutional recurrent autoencoders have been shown to perform well for unsteady flow and fluid-structure phenomena \cite{bukka2020deep,bukka2021assessment,gupta2022hybrid}.
However, the convolutional recurrent autoencoder architecture may struggle to learn PDEs with a dominant hyperbolic character and wave-like behaviour.
There are particular difficulties in constructing a low-dimensional manifold using a convolutional autoencoder and evolving these low-dimensional latent representations over a long time horizon using a recurrent neural network composed of long short-term memory cells \cite{hochreiter1997long,bukka2021assessment}.
Evolving these low-dimensional data in time with the recurrent neural network presents the challenge of having a large data set that incorporates various physical events in the training set~\cite{sorteberg2018approximating}.

The current work builds upon our previous work on the convolutional recurrent autoencoder net for the unsteady flow dynamics and fluid-structure interaction \cite{bukka2021assessment,gupta2022hybrid}. 
By employing convolutional recurrent autoencoder architecture, our objective is to develop a reduced-order model for hyperbolic PDEs and convection-dominated phenomena. 
{To reduce the dimensionality of data with dominant wave character, we investigate a denoising-based convolutional autoencoder \cite{vincent2010stacked} that can effectively compress the high dimensional data from a hyperbolic PDE.
The denoising-based convolutional autoencoder simply adds stochastic white noise to the input data. 
Following that, the model is trained to learn uncorrupted original data without noise.
The model can learn low-dimensional representations from data that includes the convective transport process.
} 
We further integrate the knowledge of numerical integration into the deep neural architecture.
We adaptively weighs the input time-steps to evolve in temporal dimension by using
an attention-based sequence-to-sequence modeling in the evolver function. 
In a nutshell, the vital contributions of the current work are as follows:
\begin{itemize}
\item Synchronous multi-time stepping prediction procedure via attention-based sequence-to-sequence modeling;
\item Assessment of a denoising-based convolutional autoencoder to learn low-dimensional manifold for hyperbolic partial differential equations;
\item Incorporation of a novel hybrid supervised-unsupervised training strategy for denoising convolutional autoencoder and evolver network simultaneously;
\item Demonstration of the proposed formulation for convection-dominated (i.e., wave propagation) physics obtained from the linear convection, nonlinear viscous Burgers, and 2D Saint-Venant shallow water equations.
\end{itemize}

The rest of the paper is organized as follows. 
Section 2 briefly presents the mathematical preliminaries. 
 The details of the proposed attention-based convolutional recurrent autoencoder net are discussed in Section 3. In Section 4, we introduce the hybrid supervised-unsupervised training strategy and explain the key aspects of the attention-based convolutional recurrent autoencoder network architecture. Section 5 presents the numerical assessment of the AB-CRAN for the one-dimensional linear convection, the viscous Burgers equation and the two-dimensional Saint-Venant shallow water problem. Finally, Section 6 concludes with a brief discussion of our findings and some directions for future research.
\section{Mathematical background}
In this section, we review the reduced-order modeling methodology, starting with a full-order numerical solution of the time-dependent parametric partial differential equation. 
The numerical solution of a parametric PDE provides the framework to generate data for a reduced-order model. 
A generic parametric partial differential equation can be presented in an abstract form:
\begin{equation}
\begin{aligned}
\pder{t}{\mathbf{U}(\mathbf{X}, t; \mathbf{\mu})}&=\mathcal{F}\left(\mathbf{U}(\mathbf{X}, t; \mathbf{\mu})\right)\quad &&\text{on} \quad \Omega \times [0, T] \times \mathcal{M},\\
\mathbf{U}(\mathbf{X},0 ; \mathbf{\mu})&=\mathbf{U}_{0}(\mathbf{\mathbf{X},\mu})\quad &&\text{on} \quad \Omega \times \mathcal{M},\\
\mathbf{U}(\mathbf{X},t ; \mathbf{\mu})&=\mathbf{U}_{\partial\Omega}(\mathbf{X},t,\mu)\quad &&\text{on} \quad {\partial\Omega} \times [0, T] \times \mathcal{M},
\label{eqn:PDE}
\end{aligned}
\end{equation}
where $\Omega \subset \mathbb{R}^{i} \ (i = 1, 2, 3)$ denotes the
spatial domain,  $\mathcal{M} \subset \mathbb{R}^{m}$ is a space of possible physical parameters for the problem, and $\mathcal{F}$ is a generic nonlinear operator describing the dynamics of the system. 
The solution field of the system is represented by $\mathbf{U}$: $\Omega \times [0, T] \times \mathcal{M} \rightarrow \mathbb{R}$ and appropriately chosen initial $\mathbf{U}_{0}(\mathbf{\mathbf{X},\mu})$ and boundary conditions $\mathbf{U}_{\partial\Omega}(\mathbf{X},{t,\mu})$. 
Here $\mu$ is the control parameters in the problem and may represent the material properties or shape parameters of interest. 
Using numerical discretization techniques (e.g., finite volume or finite element), the parametric PDE system can be discretized in the spatial domain, yielding a set of parametric nonlinear semi-discrete differential equations as follows:
\begin{equation}
\begin{aligned}
\der{t}{\mathbf{U_{N}}(\mathbf{X_{N}}, t; \mathbf{\mu})}&=\mathcal{F}_{N}\left(\mathbf{U_{N}}(\mathbf{X_{N}}, t; \mathbf{\mu})\right) &&\text{on}\ \Omega_N \times [0, T] \times \mathcal{M},\\
\mathbf{U_{N}}(\mathbf{X_{N}},0 ; \mathbf{\mu})&=\mathbf{U}_{0}(\mathbf{\mathbf{X_{N}},\mu}) &&\text{on}\ \Omega_N \times \mathcal{M},\\
\mathbf{U_{N}}(\mathbf{X_{N}},t ; \mathbf{\mu})&=\mathbf{U}_{\partial\Omega}(\mathbf{X_{N}},{t,\mu}) &&\text{on}\ {\partial\Omega_N} \times [0, T] \times \mathcal{M},
\end{aligned}
\label{eqn:discrete ODE}
\end{equation}
where $\Omega_N \subset \mathbb{R}^{N}, \ \mathbf{U_{N}}: \Omega_N \times [0, T] \times \mathcal{M} \rightarrow \mathbb{R}^{N}$ is a discrete solution and $N$ is the number of spatial degrees of freedom.  
In order to solve Eq.~(\ref{eqn:discrete ODE}), suitable time discretizations techniques are employed to evolve the spatially discretized solution in time. 
The above nonlinear semi-discrete system is common in computational physics e.g., the numerical solutions of the Euler equations and compressible Navier-Stokes equations will lead to such a nonlinear semi-discrete system. 

For given $(t ; \mathbf{\mu})$ varying in $[0, T] \times \mathcal{M}$, the set of solution fields of Eq. (\ref{eqn:PDE}) is known  as solution manifold represented by $\mathbf{S_{U}}$  as:
\begin{align}
\mathbf{S}_{\mathbf{U}}=\left[\mathbf{U}_{N,\mu_{1}}^{(t_{1})}, \ldots, \mathbf{U}_{N,{\mu}_{1}}^{({N_{T}})}, \ldots \ldots , \mathbf{U}_{N, {\mu}_{N_{\text {train }}}}^{(t_{1})}, \ldots, \mathbf{U}_{N, {\mu}_{N_{\text {train }}}}^{(N_{T})}\right].
\end{align}
When $\mathbf{\mu} \in \mathcal{M}$, the solution field of Eq. (\ref{eqn:PDE}) admits a unique solution for each $t \in[0, T]$. 
The intrinsic dimension of the solution field lying in the solution manifold is at most $n_{\mathbf{\mu}}+1 \ll N$, where $n_{\mathbf{\mu}}$ is the number of parameters. 
Herein, time essentially plays a role of additional coordinate. 
This means that each point $\mathbf{U}_{N,\mu}^{(t)}$ corresponding to $\mathbf{S_{U}}$ is completely defined in terms of at most $n_{\mathbf{\mu}}+1$ intrinsic coordinates. 
In this problem, we want to avoid solving Eq. (\ref{eqn:PDE}) by constructing a low dimensional manifold whose dimension is as close to intrinsic coordinates as possible and a time advancement strategy on this manifold exclusively from training data. Therefore, our objective is to achieve the low-dimensional approximation of the entire set of solutions to the parametric PDE given by Eq. (\ref{eqn:PDE}).
\subsection{Projection-based Reduced Order Modeling}
Projection-based reduced-order models aim to generate a low-dimensional representation that approximates the original system over a specified parameter range. 
We consider the task of finding a low-dimensional model of the system of differential equations in Eq. (\ref{eqn:discrete ODE}) with a reduced-order model as: 
\begin{align}
\der{t}{\mathbf{U}_{r}(\mathbf{X}_{r}, t; \mathbf{\mu})} = \mathcal{F}_{r}\left(\mathbf{U}_{r}(\mathbf{X}_{r},t; \mathbf{\mu})\right)\ \text{on}\ \Omega_r  
\times [0, T] \times \mathcal{M},
\end{align}
where $\mathrm{dim}(\mathbf{U}_r)<< \mathrm{dim}(\mathbf{U}_N)$. 
One approach for creating such a ROM is to introduce a reduced linear trial manifold \cite{antoulas2005approximation}. 
A linear ROM seeks to approximate the full dimension solution in the following form:
\begin{align}
\mathbf{U}_{N,\mu}^{(t)} \approx \mathbf{V} \mathbf{U}_{r,\mu}^{(t)},
\end{align}
where $ \mathbf{U}_{N,\mu}^{(t)} \in \mathbb{R}^{N}$ denotes the full state vector, the columns of the matrix $\mathbf{V} \in \mathrm{R}^{N \times r}$ contain $r$ basis vectors. 
The basis vectors in $\mathbf{V}$ are assumed to be time-invariant. 
Here $ \mathbf{U}_{r,\mu}^{(t)} \in \mathbb{R}^{r}$ denotes the low-dimensional representation often called generalized coordinates. By substituting the above subspace approximation in Eq. (\ref{eqn:discrete ODE}) will lead to:
\begin{align}
\der{t}{\mathbf{V}\mathbf{U}_{r}(\mathbf{X}_{r}, t; \mathbf{\mu})} - \mathcal{F}_{N}\left(\mathbf{V}\mathbf{U}_{r}(\mathbf{X}_{r}, t; \mathbf{\mu})\right)=\mathbf{r}(t)\ \text{on}\ \Omega_N \times [0,T] & \nonumber\\\times \mathcal{M},& 
\label{eqn:discrete reduced ODE}
\end{align}
where $\mathbf{r}(t) \in \mathbb{R}^{N}$ is the residual due to the subspace approximation \cite{schilders2008model}. 
This residual is constrained to be orthogonal to a subspace $\mathcal{W}$. 
The subspace $\mathcal{W}$ is defined by a test basis $\mathbf{W} \in \mathbb{R}^{N \times r}$, this leads to compute $\mathbf{U}_{r,\mu}^{(t)}$ such that:
\begin{align}
    \mathbf{W}^{T} \mathbf{r}(t)=0.
\end{align}
Assuming $\mathbf{W}^{T}\mathbf{V}$ is non-singular, if $\mathbf{W} = \mathbf{V}$ and $\mathbf{V}$ is orthogonal, the projection method is called a Galerkin projection \cite{volume2} and the resulting ROM can be written as:
\begin{align}
\der{t}{\mathbf{U}_{r}(\mathbf{X}_{r}, t; \mathbf{\mu})} - {(\mathbf{W}^{T} V)}^{-1}\mathbf{W}^{T} \mathcal{F}_{N}\left(\mathbf{V}\mathbf{U}_{r}(\mathbf{X}_{r}, t; \mathbf{\mu})\right) = 0\ \text{on}\ \Omega_{r} & \nonumber\\\times [0, T] \times \mathcal{M},&
\end{align}
which will yield:
\begin{align}
\der{t}{\mathbf{U}_{r}(\mathbf{X}_{r}, t; \mathbf{\mu})} - \mathbf{V}^{T} \mathcal{F}_{N}\left(\mathbf{V}\mathbf{U}_{r}(\mathbf{X}_{r}, t; \mathbf{\mu})\right) = 0\ \text{on}\ \Omega_{r} \times [0, T]& \nonumber\\\times \mathcal{M}.&
\end{align}

For constructing an efficient test basis ($\mathbf{W}$), the proper orthogonal decomposition has been widely used by the reduced-order modeling community. The POD basis vectors are computed empirically using a set of data that samples the range of relevant system dynamics. 
The POD process in conjunction with the Galerkin or Petrov-Galerkin projection techniques is commonly used to build such reduced-order models \cite{volume3}.  
One of the main disadvantages of such techniques is that they need access to the operator of the governing differential equation for evolving the basis functions in time. While the Galerkin/Petrov-Galerkin projection techniques provide interpretable and optimal reduced-order models using the known equations, they are not suitable for the development of non-intrusive data-driven models. Moreover, these projection-based ROMs pose difficulty during the construction of low-dimensional features for wave propagation and convection-dominated problems.

%
%
In order to quantify the optimality of the test subspace \cite{greif2019decay},  Kolmogorov $n$-width is used which can be stated as follows:
\begin{equation}
d_{n}(\mathcal{M}):=\inf _{\mathcal{S}_{r}} \sup _{\mathbf{U}_{N} \in \mathbf{S}_{\mathbf{U}}} \inf _{\mathbf{U}_{r} \in \mathcal{S}_{r}}\|\mathbf{U}_{N,\mu}^{(t)}-\mathbf{V}\mathbf{U}_{r,\mu}^{(t)}\|,
\end{equation}
where the first infimum is taken over all $r$-dimensional subspaces of the state space $\mathcal{S}_{r}$, and ${\mathbf{S}_{\mathbf{U}}}$ denotes the manifold of solutions over time and parameters \cite{pinkus2012n}. 
For problems governed by hyperbolic PDEs  (e.g., convection-dominated problems), the snapshot matrix exhibits slowly decaying Kolmogorov $n$-width. In such cases, the use of low-dimensional linear trial subspaces often produces inaccurate results; the ROM dimensionality must be significantly increased to yield acceptable accuracy \cite{ohlberger2015reduced}.  Indeed, the Kolmogorov $n$-width with $n$ equal to the intrinsic solution-manifold dimensionality is often quite large for such problems. 
For linear wave problems, it has been proved that the Kolmogorov $n$-width decays at a sub-exponential rate with $n$ for  non-uniform or discontinuous initial conditions \cite{greif2019decay}.
To address the $n$-width limitation of linear trial subspaces, several approaches have been pursued. In this paper, our approach involves learning a nonlinear manifold to improve its approximation properties for wave propagation and convection-dominated problems.
\section{Attention-based Convolutional Recurrent Autoencoder Net}
\subsection{Review of Convolutional Autoencoders}
To learn a nonlinear manifold, we employ a denoising-based convolutional autoencoder to train a convolutional encoder network that projects the high-dimensional data onto a low-dimensional manifold. 
Using an encoder-to-decoder network with non-linear activation functions, one can learn a mapping from a low-dimension manifold to physical space via decoders, similar to projection-based reduced order model such as proper orthogonal decomposition. 
Similar to proper orthogonal decomposition, autoencoders perform the minimization of $L_2$-norm of the error to construct the weights via backpropagation. 
Once converged,  it can be shown that the latent dimension of autoencoders span the same subspace as the proper orthogonal decomposition  \cite{baldi1989neural, plaut2018principal}.
In contrast to the POD-based reduced-order model, autoencoders can provide greater flexibility for dimensionality reduction. We use a denoising-based convolutional autoencoder as a nonlinear generalization of the POD \cite{bukka2021assessment}. 
A nonlinear ROM can be used to construct the approximation $\hat{\mathbf{U}}_{N,\mu}^{(t)}$ using the full state solutions $ \mathbf{U}_{N,\mu}^{(t)}$ as follows:
\begin{equation}
\left.
\begin{aligned}
\mathbf{U}_{r,\mu}^{(t)}&=\boldsymbol{\Psi}_{E}\left(\mathbf{U}_{N,\mu}^{(t)};\theta_{E}\right),\\
\hat{\mathbf{U}}_{N,\mu}^{(t)}&=\boldsymbol{\Psi}_{D}\left(\mathbf{U}_{r,\mu}^{(t)};\theta_{D}\right),
\end{aligned}
\hspace{1cm}
\right\}
\end{equation}
where $ \hat{\mathbf{U}}_{N,\mu}^{(t)} \in \mathbb{R}^{N}$ denotes the approximation of the full state, $\mathbf{U}_{N,\mu}^{(t)} \in \mathbb{R}^{N}$ represents the full state,  $\boldsymbol{\Psi_{E}(.;\theta_{E})}$ stands for the encoder network that maps the full state to a low-dimensional manifold, $\theta_{E}$ is the parameter of encoder, $\boldsymbol{\Psi_{D}(.;\theta_{D})}$ denotes the decoder network that maps the low-dimensional data back to the high-dimensional physical state, and $\theta_{D}$ is the parameter of decoder. 
By backpropagating the $L_2$ error, the weights of the network can be trained.
Here $ \mathbf{U}_{r,\mu}^{(t)} \in \mathbb{R}^{r}$ represents the solution on low-dimensional manifold. Our objective is to create a reduced-order model with a dimension $r$ that is as close to the intrinsic dimension $n_{\mu}+1$ of the solution manifold $\mathbf{S_{U}}$ as possible.

We employ an autoencoder as a convolutional autoencoder for the following reasons.  To begin, autoencoders need flattening multi-dimensional input data into a one-dimensional array, which can be a significant bottleneck when applied to spatial-temporal data.  Flattening leads to a loss in local spatial relationships between the data. As a result, we want the convolution filters to extract the spatially dominant structures from the multi-dimensional data.
Second, because the solution of hyperbolic PDEs is wave-like, initial disturbance propagates through the domain with a finite speed. They travel along with the characteristics of the equations.
We exploit the convolutional neural network's translational invariant property to model the hyperbolic PDE's disturbance propagation.

\begin{figure*}
    \centering
    \includegraphics[width=0.7\textwidth]{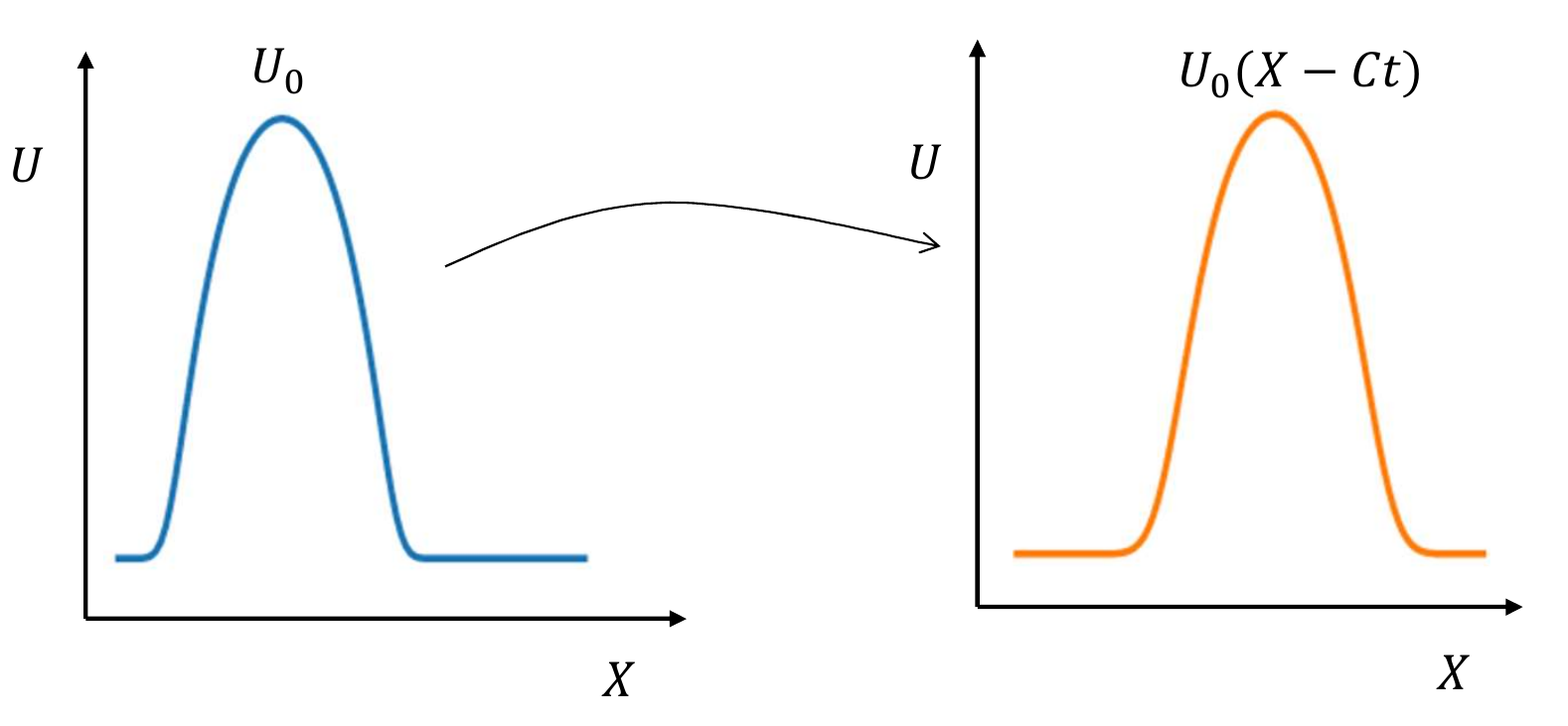}
    \caption{A simple illustration of the solution of linear convection at time $t$ by shifting initial solution $U_0$.}
\label{fig:shifting_of_solution}
\end{figure*}

To illustrate the above point further, consider a linear convection equation as a model of hyperbolic PDE:
\begin{equation}
\left.
\begin{aligned}
\frac{\partial{U}}{\partial{t}}+ C\frac{\partial{U}}{\partial{x}} &= 0, \\
U(X,0) &= U_0,
\end{aligned}
\hspace{1cm}
\right\}
\end{equation}
where $C$ denotes the wave speed. As illustrated in Fig. \ref{fig:shifting_of_solution}, the solution of the above equation is given by shifting the initial solution at time $t=0$, which can be expressed analytically at time $t$ as
\begin{equation}
  U(X,t) = U_0(X-Ct).  
\end{equation}
With regard to the data-driven prediction of this simple process, convolutional neural networks incorporate inherent translation invariance \cite{bronstein2017geometric}, making them more suitable for dealing with hyperbolic PDEs data.  
The convolutional neural network achieves translation invariance by a combination of convolutional and max-pooling layers. The convolutional layer condenses the input into a set of features and their positions.  Using the max-pooling layer, the convolutional layer's output is reduced in dimension which is accomplished by outputting only the maximum value.  As a result, the information regarding the exact location of the maximum value within the grid is discarded. This is often referred to as the translation invariance in the convolutional neural networks. The layers of convolutional neural networks are arranged into feature maps, with each unit in a feature map linked to a prior layer's local domain through a filter. 
Consider a two-dimensional input, $U \in \mathbb{R}^{N_x \times N_y}$, where a convolutional layer is composed of a collection of $F$ filters $K^f \in \mathbb{R}^{a \times b}$, each of which creates a feature map $O^f \in \mathbb{R}^{n_x \times n_y}$ through a two-dimensional discrete convolution, and non-linearity $\sigma$, which can be expressed as follows:
\begin{equation}
    \mathbf{O}_{i, j}^{f}=\sigma\left(\sum_{k} \sum_{l} \mathbf{K}_{k, l}^{f} {U}_{(i-k),(j-l)}\right).
\label{eqn:discrete_conv}
\end{equation}

\begin{figure*}
    \centering
    \includegraphics[width=\textwidth]{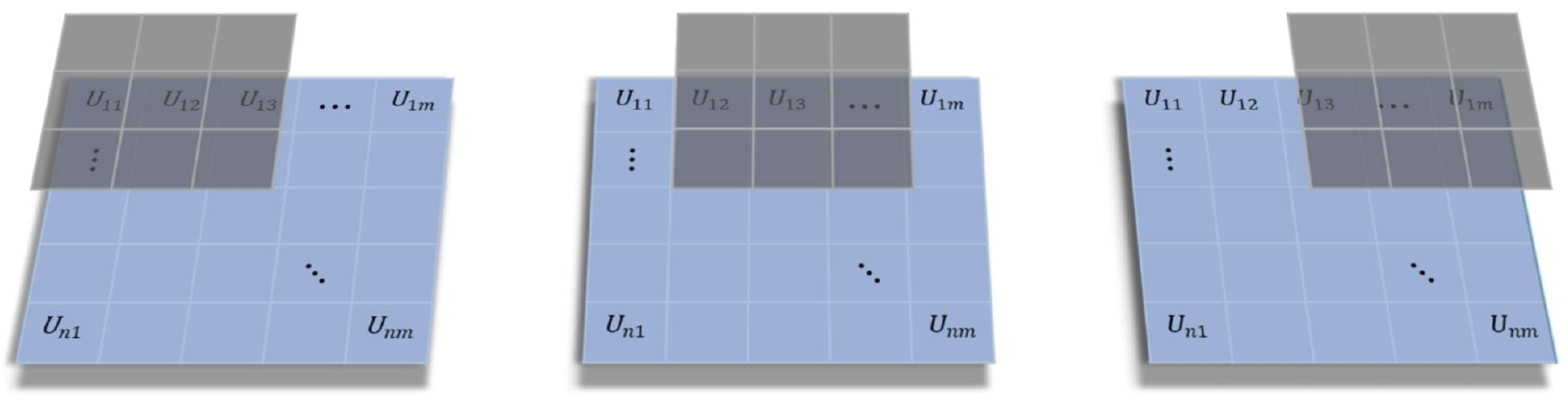}
    \caption{A simple illustration of the convolution operation with stride one and dilation rate one.}
    \label{fig:disc_conv}
\end{figure*}
\begin{figure*}
    \centering
    \includegraphics[width=\textwidth]{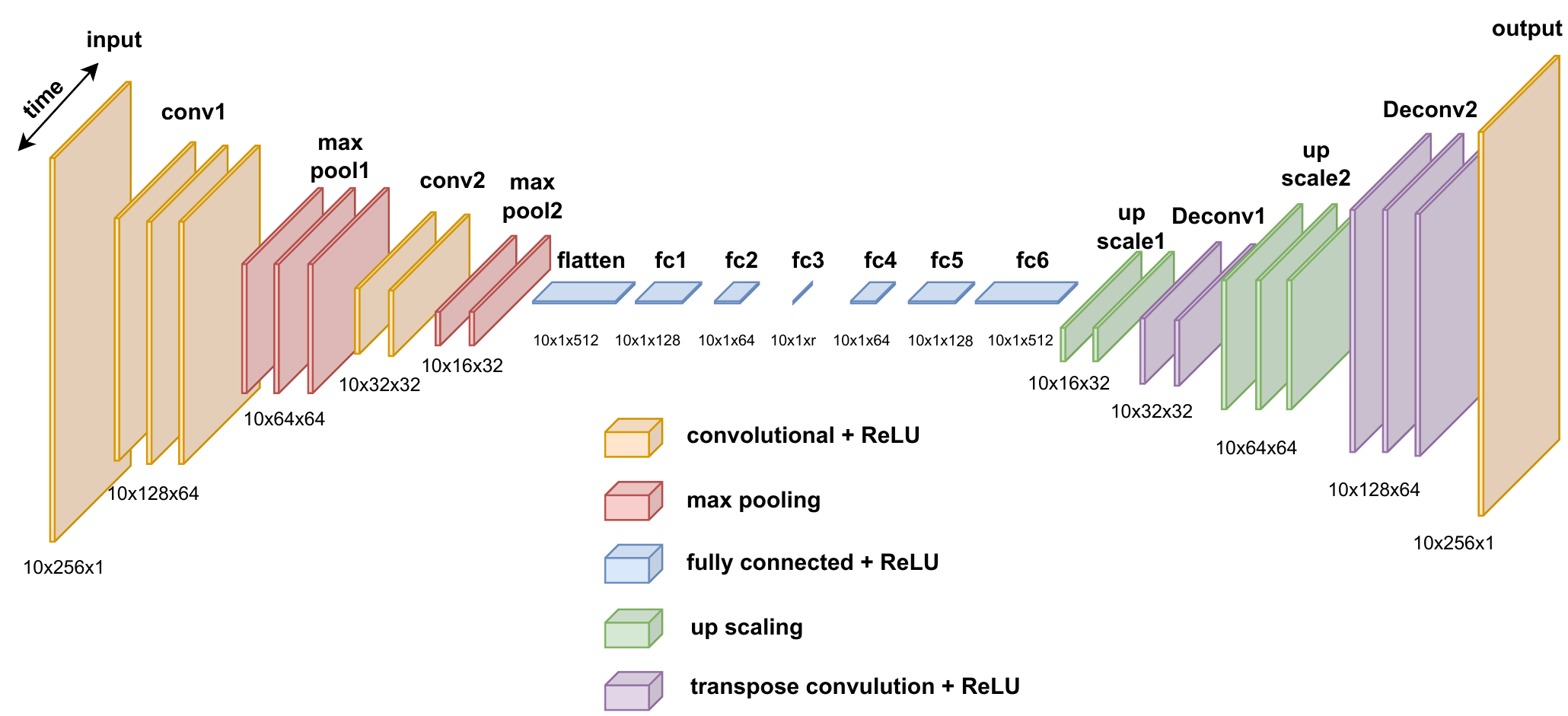}
    \caption{Schematic of convolutional autoencoder architecture employed for the linear convection problem.}
\label{fig:convAutoLin}
\end{figure*}

This operation produces a $F$-dimensional output $\mathbf{O}(\mathbf{x})=\left(o_{1}(\mathbf{x}), \ldots, o_{F}(\mathbf{x})\right)$ often referred to as the feature maps or convolutional maps. To extract the local features from a Euclidean space, the standard convolutional process can be given by: 
\begin{equation}
(\mathbf{U} \star {k})(\mathbf{x})=\int_{\Omega} \mathbf{U}\left(\mathbf{x}-\mathbf{x}^{\prime}\right) k\left(\mathbf{x}^{\prime}\right) d \mathbf{x}^{\prime}, 
\end{equation}
{where $\star$ represents convolution operator and $k$ denotes convolution function whereas the input function $\mathbf{U}$ is convolved.
For introducing point-wise non-linearity, one can employ various activation functions such as sigmoid activation $\sigma(z) = \left(1+e^{- z }\right)^{-1}$.}
Activation function allows nonlinear flow features to be captured in the discrete convolutional process. Furthermore, a pooling or down-sampling layer $\mathbf{g}=P(\mathbf{\mathbf{U}})$ may be used, given as
\begin{equation}
g_{l}(\mathbf{x})=P\left(\left\{\mathbf{U}_{l}\left(\mathbf{x}^{\prime}\right): \mathbf{x}^{\prime} \in \mathcal{N}_{e}(\mathbf{x})\right\}\right), l=1, \ldots, F,
\end{equation}
where $\mathcal{N}_{e}(\mathbf{x}) \subset \Omega_N$ is a neighborhood around $\mathbf{x}$ and $P$ is a pooling operation such as $L_{1}$, $L_{2}$ or $L_{\infty}$ norm. A convolutional network is constructed by composing several convolutional and pooling layers, obtaining a generic compositional representation as follows:
\begin{equation}
{\mathbf{\theta_{CNN}}}(\mathbf{U})=\left({K^{(L)}} \cdots P \cdots \circ {K^{(2)}} \circ {K^{(1)}}\right)(\mathbf{U})
\end{equation}
where $\mathbf{\theta}_{CNN}=\left\{K^{(1)}, \ldots, K^{(L)}\right\}$ is the hyper-vector of the network parameters consisting of all the filter banks. 
Notably, Eq.~(\ref{eqn:discrete_conv}) is modified slightly if the convolutional blocks are skipped on more than one element of the input function along any Cartesian direction. The skipping lengths along the three directions of the input is termed as the stride $s_{L}=\left[\begin{array}{ll}s_{x} \;\;  s_{y} \;\; s_{z} \end{array}\right]$ and is an important hyperparameter for the dimensionality reduction. 
Figure \ref{fig:disc_conv} {depicts discrete convolution process with a stride length of one.
Convolutional neural networks possess multi-scale characteristics which allow them to scale easily to multi-dimensional Euclidean space \cite{bronstein2017geometric}. The output features enjoy translation invariance \cite{bronstein2021geometric}, which makes CNN ideal for processing hyperbolic PDEs data.}
%
A representative sketch of the convolutional autoencoder
architecture for the linear convection problem is shown
in Fig.~\ref{fig:convAutoLin}.
%

Training this convolutional autoencoder then consists of finding the parameters that minimize the expected reconstruction error over all training examples given by
\begin{equation}
\boldsymbol{\theta}_{E}^{*}, \boldsymbol{\theta}_{D}^{*}=\arg \min _{\boldsymbol{\theta}_{\boldsymbol{E}}, \boldsymbol{\theta}_{D}}\mathcal{L}[\mathbf{U}_{N,\mu}^{(t)},\Psi_{D}(\Psi_{E}(\mathbf{U}_{N,\mu}^{(t)};\theta_{E});\theta_{D})],
\label{eqn: autoencoder loss}
\end{equation}
where $\mathcal{L}[\mathbf{U}_{N,\mu}^{(t)},\Psi_{D}(\Psi_{E}(\mathbf{U}_{N,\mu}^{(t)};\theta_{E});\theta_{D})]$ is a loss function in the $L^{2}$ norm,  which minimizes the difference between the reconstruction $\left(\Psi_{D}(\Psi_{E}(\mathbf{U}_{N,\mu}^{(t)};\theta_{E});\theta_{D})\right)$ and input $\left(\mathbf{U}_{N,\mu}^{(t)}\right)$. 
This minimization promotes $\Psi_{D} \circ \Psi_{E}$ to merely learn to copy input to output. In order to learn the latent representation of the data, under-complete autoencoders are used in which the latent dimension is less than the input dimension of the data.
There are several other regularised autoencoders that include contractive autoencoders, denoising autoencoders or sparse autoencoders, which utilize different techniques to learn robust latent representation which generalizes better for the testing data. 
Recently, the concept of self-attention in CNN layers  has been studied by  \cite{wu2021reduced} to learn global features from the fluid flow over a circular cylinder. The learning  of such global spatial features is generally suitable for parabolic PDEs but not applicable for hyperbolic PDEs and convection-dominated data.

To learn a general latent representation instead of loss in Eq.~(\ref{eqn: autoencoder loss}), a denoising autoencoder minimizes the following loss:
\begin{equation}
\mathcal{L}[\mathbf{U}_{N,\mu}^{(t)}, \Psi_{D}(\Psi_{E}(\tilde{\mathbf{U}}_{N,\mu}^{(t)};\theta_{E});\theta_{D})],
\end{equation}
where $\tilde{\mathbf{U}}_{N,\mu}^{(t)}$ is a copy of $\mathbf{U}_{N,\mu}^{(t)}$ that has been added with some form of noise. Denoising autoencoders must therefore undo this corruption rather than simply copying their input. 
Denoising training forces $\Psi_E$ and $\Psi_D$ to implicitly learn the structure of data. A denoising based convolutional-autoencoder \cite{vincent2008extracting} is explored in the current architecture which helps in discovering robust representation of the spatial-temporal data. 
As a result, the denoising based convolutional-autoencoder can be viewed as a method for defining and learning a manifold. The low dimension representation from the encoder layer, $\mathbf{U}_{r,\mu}^{(t)}=\Psi_E(\mathbf{U}_{N,\mu}^{(t)};\theta_E) $ can be thought of as a coordinate system for manifold points. 
More generally, one can think of $\mathbf{U}_{r,\mu}^{(t)}$ as a reduced-dimension representation of $\mathbf{U}_{N,\mu}^{(t)}$ which is well suited to capture the dominant variations in the data.

\subsection{Time marching via sequence-to-sequence modeling}
Next we turn our focus to the time marching problem via sequence-to-sequence modeling. To learn the system dynamics on the reduced nonlinear trial manifold in terms of the generalized coordinates, we use the mapping of the form:
\begin{equation}
\mathbf{U}^{(t+1)}_{r,\mu} = \boldsymbol{\Phi}(\mathbf{U}^{(t)}_{r,\mu};\theta_{\Phi}).
\end{equation}
Using deep learning, one can address the difficulty of the evolution of generalized coordinates in the projection-based model reduction where access to operators in the governing laws is needed to evolve the basis functions.

Recurrent neural networks have traditionally been used in the deep learning community to model the time evolution of a state variable. 
At each step, the recurrent neural networks compute an $m$-dimensional summary vector $\mathbf{h(t)}$ of all input steps up to and including $t$. 
This partial summary is computed using a shared update function, $\mathcal{R}: \mathbb{R}^{r} \times \mathbb{R}^{m} \rightarrow \mathbb{R}^{m}$, based on the current step's features and the previous step's summary as follows:
\begin{equation}
\mathbf{h}^{(t)}=\mathcal{R}\left(\mathbf{U}^{(t)}_{r,\mu}, \mathbf{h}^{(t-1)}\right).
\end{equation}
In a simple recurrent neural network model \cite{elman1990finding}, one can consider $\mathbf{U}^{(t)}_{r,\mu}$ and $\mathbf{h}^{(t-1)}$ as flat vector representation and $\mathcal{R}$ can be expressed as a single fully-connected neural network layer which can be written as:
\begin{equation}
\mathbf{h}^{(t)}=\sigma\left(\mathbf{W} \mathbf{U}^{(t)}_{r,\mu}+\mathbf{Q h}^{(t-1)}+\mathbf{b}\right),
\end{equation}
where $\mathbf{W} \in \mathbb{R}^{m \times r}, \mathbf{Q} \in \mathbb{R}^{m \times m}$ and $\mathbf{b} \in \mathbb{R}^{m}$ are learnable parameters and
$\sigma$ is  a non-linear activation function. 
The summary vectors can then be used appropriately for the downstream task whenever a prediction is needed at each step of the sequence. Subsequently, a shared predictor can be applied to each $\mathbf{h}^{(t)}$ individually. 
The initial summary vector, in particular, is usually either set to a zero-vector, i.e. $\mathbf{h}^{(0)}=\mathbf{0}$, or made learnable. 
To address the issues in time-series modeling such as long-term dependency in the input data, and vanishing gradients, long short-term memory cells are utilized in the present work. 
Learning the system dynamics of hyperbolic PDEs with a long short-term memory cell evolver is difficult to generalize for unforeseen input conditions and predict outputs for large time horizons.

\begin{figure*}
    \centering
    \includegraphics[width=0.9\textwidth]{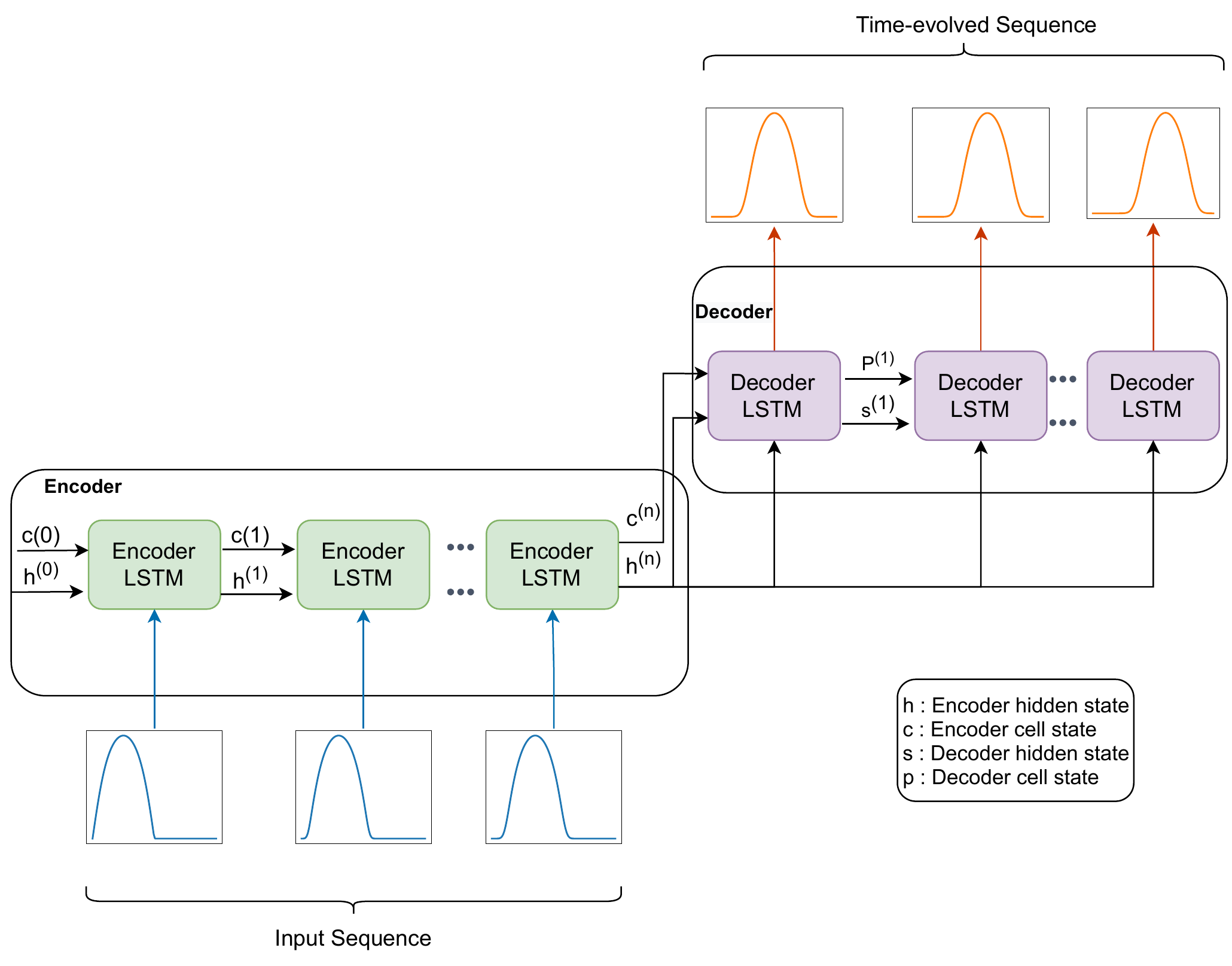}
    \caption{ {Illustration of sequence-to-sequence long short-term memory cell evolver. Using the input sequence, the encoder long short-term memory block transforms and provides the complete history of input sequence into the final  hidden state $h^{(n)}$. The final cell state $c^{(n)}$ and the hidden state  $h^{(n)}$ are used as input for the decoder's computation and the decoder iterates over the hidden states to generate the time evolved sequence.
    }}
\label{fig:seq2seq}
\end{figure*}

In deep learning, sequence-to-sequence modeling has recently seen widespread application in sequential data processing and natural language processing~\cite{cho2014learning}. 
Sequence-to-sequence architecture is a general end-to-end approach for learning sequence data that makes few assumptions about the sequence structure.  
It also often employs an encoder-decoder structure to encode the input sequence to a fixed-dimensionality vector and then decode the target sequence from the vector~\cite{sutskever2014sequence}. 
A stack of long short-term memory cells serves as the encoder. The encoder summarises the information into a  context vector. The final states produced by the model's encoder are utilized in initializing the decoder stack. 
The context vector is used as an input to the decoder stack to generate the output sequence synchronously. 
A stack of long short-term memory cells also makes up the decoder.
Each long short-term memory cell in the decoder takes a hidden state and cell state from the previous cell and the context vector as input and produces the output.
The entire sequence-to-sequence procedure via long short-term memory evolver is depicted in  Fig. \ref{fig:seq2seq}. Next, we present the attention mechanism which aims to provide an improvement to the encoder-decoder architecture.
\subsubsection{Temporal attention mechanism}
The aim of a data-driven learning model for multi-step time series prediction is to implement a mapping from a sequence of input data, $(\mathbf{U}_{r,\mu}^{(1)}, \mathbf{U}_{r,\mu}^{(2)},\dots, \mathbf{U}_{r,\mu}^{(t)} )$, to an output sequence:
\begin{align}
\{\mathbf{U}_{r,\mu}^{(t+1)},\mathbf{U}_{r,\mu}^{(t+2)},\dots,\mathbf{U}_{r,\mu}^{(2t)}\}=\Phi\left(\mathbf{U}_{r,\mu}^{(1)}, \mathbf{U}_{r,\mu}^{(2)},\dots, \mathbf{U}_{r,\mu}^{(t)};\theta_{\Phi}\right).
\end{align}
Using a collection of training data and respective labels, the model $\Phi$ is usually estimated through supervised learning with a direct strategy for multi-step prediction.
While the sequence-to-sequence model works reasonably well for a short sequence, it becomes increasingly difficult to summarize a long sequence of vectors into a single vector. 
When the length of the sequence grows, the model frequently forgets the earlier parts of the input sequence when processing the last parts. 
An attention mechanism can deal with this problem.  
An attention layer assigns proper weight to each hidden state output from the encoder and maps them to the output sequence. In a loose sense, attention attempts to mimic cognitive attention in neural networks. In the encoder-decoder architecture, the key role of attention is to facilitate the decoder to selectively access the encoder information during decoding.

\begin{figure*}
    \centering
    \includegraphics[width=\textwidth]{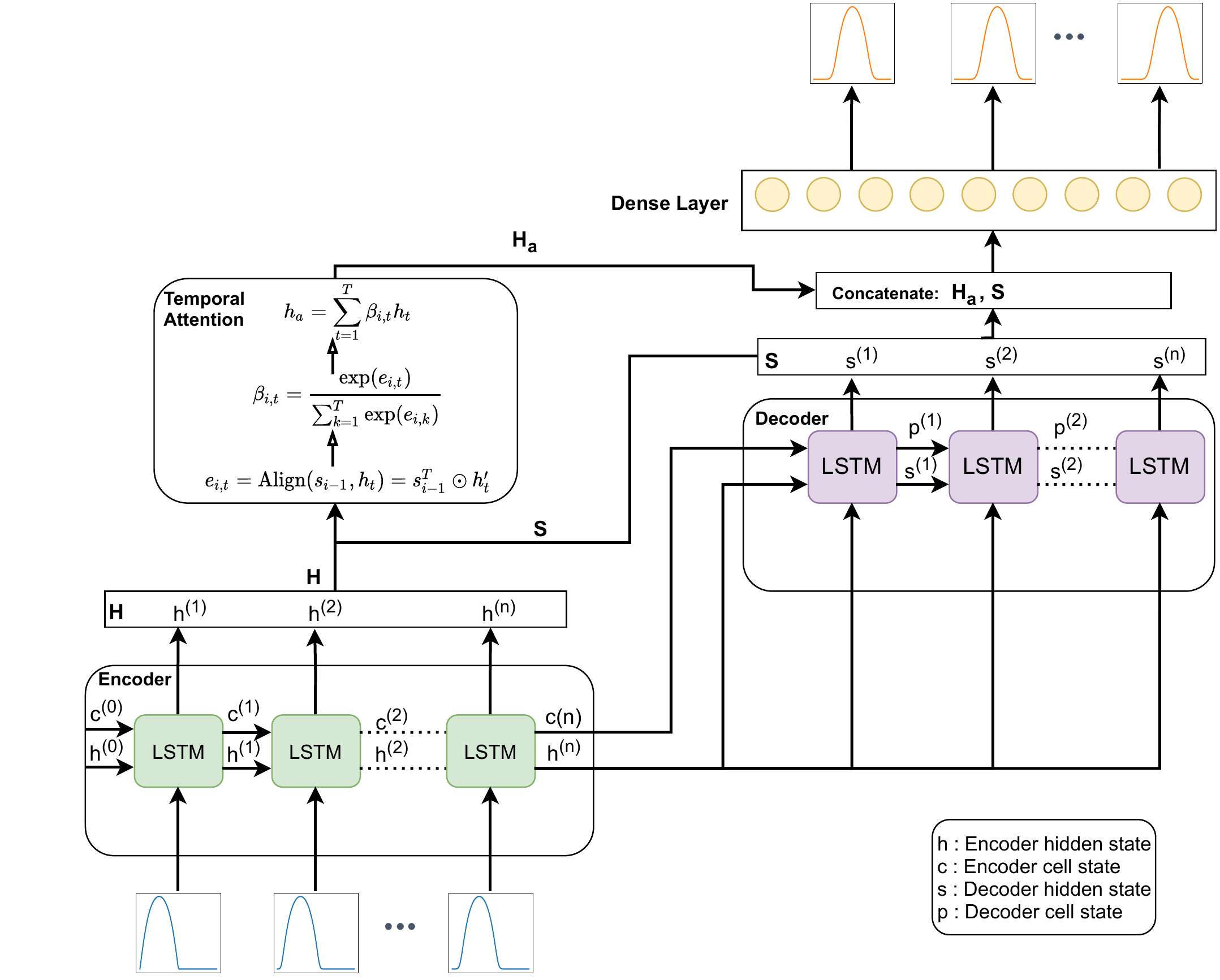}
    \caption{{Illustration of the proposed attention-based sequence-to-sequence evolver. While the encoder generates hidden state vectors $H$ by transforming input, the decoder generates hidden state ($S$) by iterating over final encoder hidden state $h^{(n)}$. Notably the alignment score between $H$ and $S$ are computed.}}
\label{fig:attentionSeq2Seq}
\end{figure*}

For our sequence-to-sequence model, a complete process of the temporal attention mechanism is demonstrated in Fig. \ref{fig:attentionSeq2Seq}. The interface between the encoder and the decoder is constructed and marked as a temporal attention layer. 
%
The context vector represents the initial hidden state for the decoder by encapsulating the information of the input elements.
We employ a long short-term memory cell as the encoder, which can take a time-series sequence $\mathbf{U}_{r,\mu} = (\mathbf{U}_{r,\mu}^{(1)}, \mathbf{U}_{r,\mu}^{(2)},\dots, \mathbf{U}_{r,\mu}^{(t)} )$ as the input data and process it recursively while maintaining its internal hidden states $\mathbf{h}^{(t)}$. At each time step t, the long short-term memory cell reads $\mathbf{U}_{r,\mu}^{(t)}$ and updates its hidden state $\mathbf{h}^{(t)}$ as follows:
\begin{align}
{h}^{(t)}=\mathrm{L S T M}\left(\mathbf{U}_{r,\mu}^{(t)}, \mathbf{h}^{(t-1)}, {c}^{(t-1)}\right),
\end{align}
where $\mathbf{h}^{(t-1)}$ is hidden state at time $t-1$ and ${c}^{(t-1)}$ is cell state at $t-1$.
Subsequently, as a weighted sum of the encoder network's hidden states, the temporal attention context vectors are formed, which are being used to identify the encoder's hidden representation and redirect the decoder to attend to these hidden states.  

The temporal attention layer computing process can be given by:
\begin{equation}
\left.
\begin{aligned}
e_{i, t}&=\mathbf{S}^{(t)} \odot \mathbf{H}^{(t)}, \\
\beta_{i, t}&=\frac{\exp \left(e_{i, t}\right)}{\sum_{k=1}^{T} \exp \left(e_{i, k}\right)}, \\ 
h_{a}&=\sum_{t=1}^{T} \beta_{i, t} \mathbf{h}^{(t)},
\end{aligned}
\hspace{1cm}
\right\}
\end{equation}
$e_{i, t}$ represents the soft align computation between the hidden state $s^{(i)}$ of the decoder layer and the hidden state $\mathbf{h}^{(t)}$ of the encoder layer.
Herein, $\beta_{i, t}$ indicates the attention weights that correspond to the importance of the input time series frame at time-step $t$ to predict the output value at time-step $i$, which employs the soft-max function to normalize the vector $e_i$ of length $T$ as the attention mask over the input time series sequence. The variable $h_a$  is the final state of the attention layer. 
Further details about the attention mechanism can be found in Vaswani et al. \cite{vaswani2017attention}.
If only the initial condition is given instead of the input sequence of data, our AB-CRAN algorithm can still be applied to predict the spatial-temporal history with a minor modification.
The framework can be applied $n_t$ times to a single time-step auto-regressively to generate the first $n_t$ length sequence, and then the method can be applied in a format of sequence-to-sequence to generate the entire spatial-temporal history. 

\subsubsection{Time integration analogous in AB-CRAN}
\begin{figure*}
    \centering
    \includegraphics[width=\textwidth]{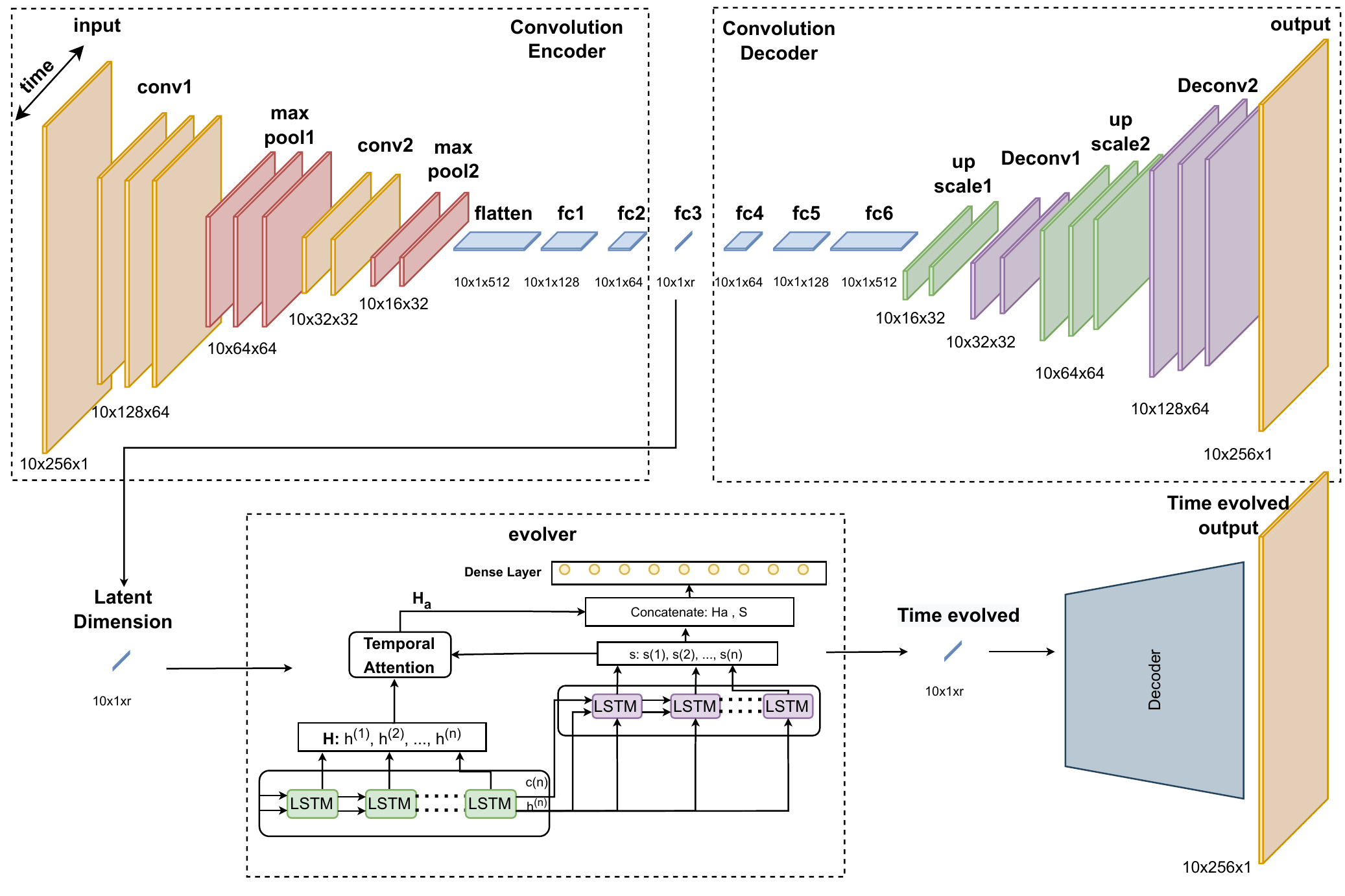}
    \caption{{Visualization of attention-based convolutional recurrent autoencoder architecture.  Three blocks are shown namely the convolution encoder for creating the latent low-dimensional representation, the evolver for propagating the low-dimensional feature in time and the decoder for transforming the low dimension space to input data space.}}
\label{fig:attention_CRAN_Lin}
\end{figure*}

To emulate the behavior of hyperbolic PDEs, an architecture must be adaptable enough to capture both the temporal evolution of the initial disturbance and the spatial variation.  For this purpose, we develop an attention-based sequence-to-sequence long short term memory cells for evolving latent dimensions in time over a long time horizon. The state-of-the-art numerical methods such as Euler-Forward difference in time utilize information from the neighboring cells at a time level $n$ with $\mathbf{U}_{i-l}^{n},\ldots,\mathbf{U}_{i}^{n},\ldots,\mathbf{U}_{i+l}^{n}$ to calculate the solution $\mathbf{U}_{i}^{n+1}$ at time level $n+1$.
To illustrate this, let us consider a semi-discretized form of the differential system:
\begin{align}
\frac{d \mathbf{U}_{i}}{d t}=f_{i}\left(\mathbf{U}_{0}, \mathbf{U}_{1}, \ldots, \mathbf{U}_{n_{x-1}}\right), \quad i=0, \ldots, n_{x-1}.
\end{align}
To estimate the value at time level $n+1$, the forward Euler time integration can be written as:
\begin{align}
\mathbf{U}_{i}^{n+1}=\mathbf{U}_{i}^{n}+dtf_{i}\left(\mathbf{U}_{0}, \mathbf{U}_{1}, \ldots, \mathbf{U}_{n_{x-1}}\right), \quad i=0, \ldots, n_{x-1}.
\end{align}
Our architecture utilizes the entire spatial grid and extracts the dominant features using the feature maps from the convolutional layer. After that, the architecture projects the data onto a low dimensional manifold via denoising-based convolution autoencoder $\mathbf{U}_{r,\mu}^{(t)}=\boldsymbol{\Psi}_{E}\left(\mathbf{U}_{N,\mu}^{(t)};\theta_{E}\right)$. 
The input sequence of data is transformed to the generalized coordinates and the attention mechanism is employed to weigh the generalized coordinates and create a summary vector of the entire sequence. To predict solution at multiple time levels ($\mathbf{U}^t,\ldots,\mathbf{U}^{t+k}$), the encoded context vector is passed through the decoder long short-term memory cell which allows to capture the propagation of disturbance over a time horizon:
\begin{equation}
\left.
\begin{aligned}
\left\{{h}^{(t)},\dots,{h}^{(t+k)}\right\}&=\mathrm{L S T M}\left(\mathbf{U}_{r,\mu}^{(t)},\dots,\mathbf{U}_{r,\mu}^{(t+k)}, {h}^{(0)}, {c}^{(0)}\right),\\
h_{a}&=\sum_{t=1}^{k} \beta_{i, t} h^{(t)},
\end{aligned}
\right\}
\end{equation}
where $\beta$ is a weighting coefficient for different input time-steps. To evolve the reduced latent features, the LSTM neural networks are used for the temporal evolution. 
The proposed novel attention-based convolutional recurrent autoencoder incorporates the implicit biases required to solve hyperbolic PDEs in the network architecture. 
The denoising-based convolutional autoencoder takes advantage of translational invariance to capture the shifting of the initial solution in all spatial locations. The attention-based sequence-to-sequence long short term memory cells can encode the input sequence and predict multiple time steps in the future. Figure \ref{fig:attention_CRAN_Lin} shows a representative architecture of the proposed AB-CRAN  for the linear convection problem.


\section{Training strategy for AB-CRAN}
In the present work, the training of the denoising convolutional autoencoder together with the attention-based sequence-to-sequence LSTM network is a crucial component. The main challenge lies in preventing either the convolutional autoencoder or the evolver of the model from over-fitting. We further elaborate on the training dataset construction and the training process.

\begin{figure*}
  \centering
\includegraphics[width=0.8\textwidth]{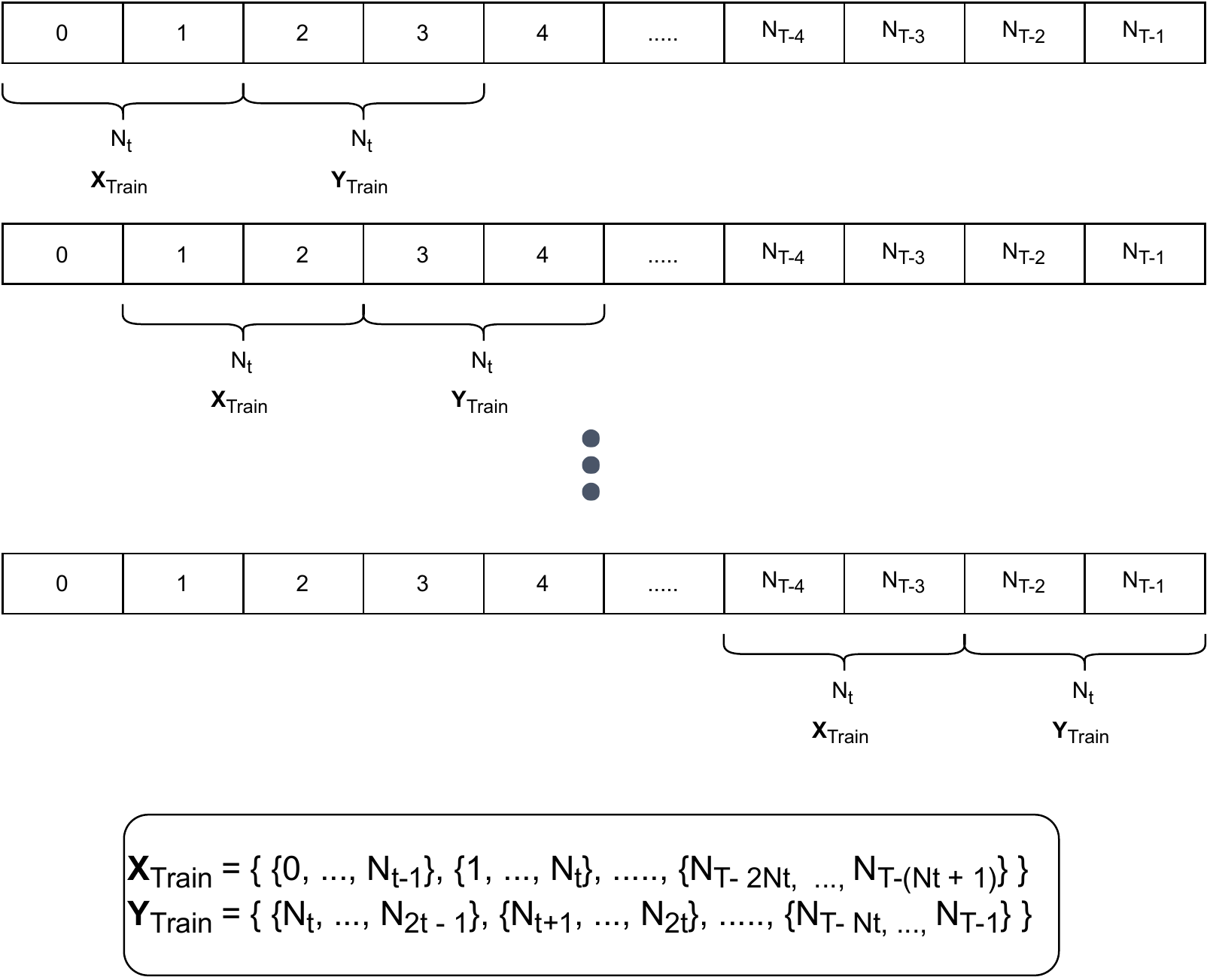}
  \caption{An illustration of training set generation from high-dimensional data via the AB-CRAN procedure.}
  \label{fig:train_set}
\end{figure*}

Consider the following data notation: $\mathcal{U} = \left\{\mathbf{U}_{N,\mu 1}^{(1)},\ldots, \mathbf{U}_{N,\mu 1}^{(N_T)}, \ldots,\mathbf{U}_{N,\mu_{N}}^{(1)},\ldots, \mathbf{U}_{N,\mu_{N}}^{(N_T)} \right\} \in \mathbb{R}^{N_\mu \times N_T \times N}$, where $N_{\mu}$ is the number of physical parameters, $N_T$ is number of time-steps, and $N$ is the spatial dimension. 
For a specific parameter  $\mu_i$, the spatial-temporal data $\mathbf{U}_{N,\mu i} \in \mathbb{R}^{N_T \times N}$ is referred to as a snapshot matrix of the parametric partial differential equation.
For the sequence-to-sequence prediction, the snapshot matrix is further divided into two matrices $\mathbf{X}_{\text{Train}}\text{ and } \mathbf{Y}_{\text{Train}}$, of which $\mathbf{X}_{\text{Train}}$ serves as an input and $\mathbf{Y}_{\text{Train}}$ as the ground truth for the AB-CRAN predictions. Each snapshot matrix of parametric PDE with specific parameter is divided into $N_{s}$ sets, where $N_{s} = N_{T} - 2N_{t} + 1$, and each subset contains 2$N_t$ time-steps length.
Here, $N_t$ denotes the length of the input sequence for the evolver function.  
The first $N_t$ time-steps of each $N_s$ subsets are taken to form $\mathbf{X}_{\text{Train}}$ and the last $N_t$ time-steps from each $N_s$ subsets form the $\mathbf{Y}_{\text{Train}}$.
The shape of $\mathbf{X}_{\text{Train}}\text{ and } \mathbf{Y}_{\text{Train}}$ is $N_{\mu} \times N_s \times N_t \times N$. 
The entire procedure for creating $\mathbf{X}_{\text{Train}}\text{ and } \mathbf{Y}_{\text{Train}}$ matrices is depicted in Fig.~\ref{fig:train_set}. 
Finally, for training $\mathbf{X}_{\text{Train}}\text{ and } \mathbf{Y}_{\text{Train}}$ matrices are reshaped into $N_m \times N_t \times N$ with $N_m = N_{\mu} \times N_s$.

For an improved neural network training and to prevent the over-saturation of any particular feature, the training data are scaled appropriately as follows: 
\begin{equation}
\left.
\begin{aligned}
\overline{X}_{\text{Train}}&=\frac{\overline{X}_{\text{Train}} - \overline{X}_{\text{Train}_{,\min }}}{\overline{X}_{\text{Train}_{,\max }}-\overline{X}_{\text{Train}_{,\min }}},\\
\overline{Y}_{\text{Train}}&=\frac{\overline{Y}_{\text{Train}} - \overline{Y}_{\text{Train}_{,\min }}}{\overline{Y}_{\text{Train}_{,\max }}-\overline{Y}_{\text{Train}_{,\min }}},
\end{aligned}
\hspace{1cm}
\right\}
\end{equation}
where $\overline{X}_{\text{Train}}, \overline{Y}_{\text{Train}}  \in[0,1]^{N_m\times N_t \times N}$ is a min-max normalized training data. 
To train the autoencoder and the evolver parts of the network simultaneously, a hybrid supervised-unsupervised training strategy is devised in the present work.  
For the unsupervised training of the denoising convolutional autoencoder, the training dataset is added with Gaussian noise: $ \tilde{X}_{\text{Train}} = \overline{X}_{\text{Train}} + \mathcal{N}(mean,SD) \in \mathbb{R}^{N_{m} \times N_{t} \times N} $. 
The autoencoder loss is computed by comparing reconstruction ($\Psi_{D}(\Psi_{E}(\tilde{X}_{\text{Train}};\theta_E);\theta_D)$) with the uncorrupted normalized data input $\overline{X}_{\text{Train}}$. 
For the supervised training of evolver, to emulate the temporal evolution, the network output is compared with the ground truth $\mathcal{Y} \in \mathbb{R}^{N_m \times N_t \times N}$ where each sequence is shifted in time by $N_t$ time-steps with respect to the input sequence. 

To train both the components of the convolutional recurrent autoencoder model, our strategy is to split the forward pass into two stages. 
In the first stage, the encoder takes an $N_{b}$-sized batch of the Gaussian training data $\tilde{X}_{\text{Train}}^{b} \subset \tilde{X}_{\text{Train}}$, where $\tilde{X}_{\text{Train}}^{b} \in [0,1]^{N_{b} \times N_{t} \times N}$, and outputs the current $N_{b}$-sized batch of low-dimensional representations of the training sequence (${H}^{b} \in \mathbb{R}^{N_b \times N_t \times r}$).
{The decoder takes this low-dimensional representation and builds a reconstruction function ($\hat{{X}}^{b} \in \mathbb{R}^{N_b \times N_t \times N}$).} 
{The normalized input without the Gaussian noise, represented as $\overline{X}_{\text{Train}}^{b}$, is compared with the reconstruction function which is employed for the autoencoder loss.}  
In the second stage of the forward pass, we evolve the low-dimensional representation (${H}^{b}$) by passing through the attention-based sequence-to-sequence model via long short-term memory cell. 
To obtain the low-dimensional representation in the physical domain, the evolved low-dimensional representation is transmitted through a decoder network to recover the physical dimensions. 
The evolved physical dimension { (${X}^{\prime b} \in \mathbb{R}^{N_b \times N_t \times N}$)} is compared with the ground truth to form the evolver loss. 
{We seek to construct a loss function that weights the error during the full-state reconstruction. In that way, the evolution of the low-dimensional representations provides us improved control over the nonlinear manifold learning and the evolver part of the network. 
For this purpose, we introduce a hyper-parameter $\alpha$, which combines the autoencoder loss and the evolver loss into a single loss value. }

To find the model parameters  $\theta^* = \left\{\theta_E^*,\theta_D^*,\theta_\Phi^*\right\}$ such that for any sequence $\mathbf{U}_{\mu}=$ $\left[\mathbf{U}_{N,\mu}^{1}, \ldots, \mathbf{U}_{N,\mu}^{N_{T}}\right]$ the expected error between the model and the data can be minimized as follows:
\begin{equation}
\begin{aligned}
\mathcal{J}(\boldsymbol{\theta}) =\mathcal{L}\left(\mathcal{\overline{U}}, \tilde{\mathcal{U}} \mathcal{Y}\right)
=\frac{1}{N_m}\sum_{j=1}^{N_m}\left( \frac{\alpha}{N_t}\sum_{i=1}^{N_{T}} \left\|\hat{U}_{N, \mu_j}^{i}-{U}_{N, \mu_j}^{i}\right\|_{2}^{2} \right. \\\left.+ \frac{ (1-\alpha)}{N_T}\sum_{i=1}^{N_{t}} \left\|{U}_{N, \mu_j}^{\prime i}-{Y}_{N, \mu_j}^{i}\right\|_{2}^{2}\right),
\label{eqn:loss}
\end{aligned}
\end{equation}
here $\mathcal{\overline{U}},\tilde{\mathcal{U}}, \mathcal{Y}$ denote the normalized input data matrix, the Gaussian data matrix, and the ground truth matrix, respectively. $\hat{U}_{N, \mu_j}^{i}$ denotes the autoencoder reconstruction from the model and ${U}_{N, \mu_j}^{\prime i}$ represents time-evolved prediction from the evolver.
At every training step, the autoencoder performs a regular forward pass while constructing a new batch of low-dimensional representation. The low-dimensional representations are subsequently used to train the evolver.  In the present work, we utilize a step-based learning rate scheduler which reduces the initial learning rate by a decay factor on every predetermined number of steps \cite{bengio2012practical}.
Twenty percent of the training data are chosen and set aside for the purpose of validation. The validation data set is being used for the purpose of early stopping \cite{yao2007early}. We use the ADAM optimizer \cite{kingma2017adam}, a version of stochastic gradient descent that employs momentum for updating the parameters.
Algorithm \ref{alg:alg1} provides a complete training procedure for our AB-CRAN architecture.
  

\SetKwInput{KwOutput}{Output}

\begin{algorithm}
\caption{AB-CRAN Training Algorithm}
\KwIn{$\tilde{X}_{\text{Train}}$, $\overline{X}_{\text{Train}}$, $\overline{Y}_{\text{Train}}$, $N_{\text{epochs}}$, $N_{b}$, $\alpha$}
\KwOutput{$\theta^* = \left\{\theta_E^*,\theta_D^*,\theta_\Phi^*\right\}$}
 Initialize $\theta$\;
 \While{ epoch $<$ $N_{\text{epochs}}$ }
 {
 Randomly sample batch from training data: $\tilde{X}_{\text{Train}}^{b} \subset \tilde{X}_{\text{Train}}$\;
 Encoder forward pass: ${\mathcal{H}}^{b} \leftarrow \Psi_{E}\left(\mathcal{\tilde{X}}^{b_{A E}};\theta_{E}\right)$, ${\mathcal{H}}^{b} \in \mathbb{R}^{N_b \times N_t \times r}$\;
 Decoder forward pass: $\hat{\mathcal{X}}^{b_{A E}} \leftarrow \Psi_{D}\left({\mathcal{H}}^{b};\theta_{D}\right)$, $\hat{\mathcal{X}}^{b_{A E}} \in \mathbb{R}^{N_b \times N_t \times N}$\;
 Evolver forward pass: $\mathcal{H}^{\prime b} \leftarrow \Phi\left({\mathcal{H}}^{b};\theta_{\Phi}\right)$, $\mathcal{H}^{\prime b} \in \mathbb{R}^{N_b \times N_t \times r}$\;
Evolved physical space: $\mathcal{X}^{\prime b} \leftarrow \Psi_{D}\left(\mathcal{H}^{\prime b};\theta_{D}\right)$, $\mathcal{X}^{\prime b} \in \mathbb{R}^{N_b \times N_t \times N}$\;
Calculate loss $\mathcal{L}$ via Eq.~(\ref{eqn:loss})\;
Estimate gradients $\hat{\mathbf{g}}$ using automatic differentiation\;
Update parameters: $\theta \leftarrow A D A M(\hat{\mathbf{g}})$\;
}
Updated parameters: $\{\theta^{*}  = \theta_{E}^{*},\theta_{D}^{*},\theta_{\Phi}^{*}\}$
\label{alg:alg1}
\end{algorithm}

\begin{algorithm}
\caption{AB-CRAN Prediction Algorithm}
\KwIn{$\overline{X}_{\text{in}}$, $N_{th}$}
\KwResult{Model prediction $\hat{X}_{\text{out}}$ }
 Load trained parameter $\theta^{*} =\{\theta_{E}^{*},\theta_{\Phi}^{*},\theta_{D}^{*}\}$\;
 Encoder forward pass: ${\mathcal{H}} \leftarrow \Psi_{E}\left(\mathcal{X}^{b_{A E}};\theta_{E}^{*}\right)$\;
 \While{ i $<$ $N_{\text{th}}$ }
 {
 Evolver forward pass: $\mathcal{H}^{\prime} \leftarrow \Phi\left({\mathcal{H}};\theta_{\Phi}^{*}\right)$\;
Evolved physical space: $\mathcal{X}^{\prime} \leftarrow \Psi_{D}\left(\mathcal{H}^{\prime};\theta_{D}^{*}\right)$\;
Append: $\hat{X}_{\text{out}} \leftarrow \mathcal{X}^{\prime}$\;
}

Output: $\hat{X}_{\text{out}}$
\label{alg:alg2}
\end{algorithm}
The process of prediction becomes relatively simple once the model has been trained. The encoder network is used to generate a low-dimensional representation of the input sequence $\overline{X}_{\text{in}}$ using the trained parameters $\theta^{*}$. This low-dimensional representation ($\tilde{\mathcal{H}}$) is then evolved for $n$ time-horizons ($N_{th}$) by iterative application of evolver network. The user can rebuild the full-dimensional state from $\tilde{\mathcal{H}}$ at any time horizon using the decoder part of the AB-CRAN framework. Algorithm \ref{alg:alg2} explains how to make a prediction using our proposed AB-CRAN framework.


\section{Numerical Results}
In this section, we show how the proposed architecture can predict the evolutionary behavior of hyperbolic PDEs. 
The effectiveness of the proposed methodology will be demonstrated by solving the 1D linear convection equation, the 1D nonlinear viscous Burgers equation and the 2D Saint-Venant shallow water system.

\subsection{Linear convection equation}
As a first test case, let us consider the linear convection equation, whose solution $U$ in the domain $\Omega \equiv [0, 1]$ satisfies the parametric partial differential equation given by:
\begin{align}
 \frac{\partial U}{\partial t}+\mu \frac{\partial U}{\partial X}=0 \quad \text{in~} \quad\Omega, 
\end{align}
with the following initial condition:
\begin{align}
   U(x, 0)=U_{0}(x) \equiv f(x),
\end{align}
where $\mu \in [0.775,1.25]$ denotes the wave phase speed. Here, $ f(x)=(1 / \sqrt{2 \pi \rho}) e^{-x^{2} / 2 \rho}$, and we set $\rho=10^{-4}$.
{The parameters have been chosen at random and without bias to represent the most general cases, such as $\sigma = 10^{-4}$; however, any other value can be used without forgoing generality.}
The exact solution is simply $U(x, t)=f(x-\mu t)$ which is used to generate the ground truth data. The dataset is built by using the exact solution in the space-time domain $[0, L] \times [0, T]$, by setting $L=1$ and $T = 1$.
In this problem, a one-dimensional spatial discretization of 256 grid nodes and 200 time steps are used. We consider $N_{\mu}=20$ training-parameter instances uniformly distributed over $\boldsymbol{\mu}$ and $N_{\text {test }}=19$ testing-parameter instances such that $\mu_{\text {test }, i}=\left(\mu_{\text {train }, i}+\mu_{\text {train }, i+1}\right) / 2$, for $i=1, \ldots, N_{\text {test }}$.

The details of the architecture of this test case are as follows. We choose a 15-layers  AB-CRAN net. The encoder consists of a convolutional layer, maxpooling and fully connected layers. There are {$r$} neurons in the output layer of the encoder function, where {$r$} corresponds to the dimension of the reduced trial manifold. Specific details of the encoder, the decoder and the evolver functions are summarized in Table \ref{tab:LC_ENC}, \ref{tab:LC_prop} and \ref{tab:LC_Dec}. The total number of trainable parameters (i.e., weights and biases) of the neural network for this case is 199,461. 
{The model was trained from scratch with TensorFlow \cite{tensorflow2015-whitepaper} using single NVIDIA P100 Pascal graphical processing unit, and training converges in 285 epochs and 10 minutes of wall clock time.}

\begin{table*}
\caption{Attributes of convolutional and dense layers in the encoder $\boldsymbol{\Psi}_{E}(.;\theta_E)$.}
\centering
\begin{ruledtabular}
\begin{tabular}{lllllll}
\text { Layer } & $\begin{array}{l}
\text { Layer } \\
\text { Type }
\end{array}$& $\begin{array}{l}
\text { Input } \\
\text { Dimension }
\end{array} $& $\begin{array}{l}
\text { Output } \\
\text { Dimension }
\end{array} $& $\begin{array}{l}
\text { Kernel } \\
\text { Size }
\end{array} $& $\begin{array}{l}
\text { \# filters/ } \\
\text { \# neurons }
\end{array} $& \text { Stride }\\
\hline 1 & \text{Conv 1D} & {[10,256,1]} & {[10,128,64]} & {[5]} & 64 & 2 \\
 & \text{MaxPool 1D} &{[10,128,64]} & {[10, 64, 64]} & - & - & - \\
2 & \text{Conv 1D} & {[10,64,64]} & {[10,32,32]} & {[5]} & 32 & 2 \\
 & \text{MaxPool 1D} &{[10,32,32]} & {[10, 16, 32]} & - & - & - \\
 & \text{Flatten} &{[10,16,32]} & {[10, 512]} & - & - & - \\
3 & \text{Dense} &{[10,512]} & {[10,128]} & -  & 128 & -\\
4 & \text{Dense} &{[10,128]} & {[10,64]} & -  & 64 & -\\
5 & \text{Dense} & {[10,64]} & {[10,r]} & -& r&- \\
\end{tabular}

  \label{tab:LC_ENC}
\end{ruledtabular}
\end{table*}

\begin{table*}
\caption{Attributes of evolver functions $\boldsymbol{\Phi}(.;\theta)$.}
\centering
\begin{ruledtabular}

\begin{tabular}{lllllll}
\text { Layer } & $\begin{array}{l}
\text { Layer } \\
\text { Type }
\end{array}$& $\begin{array}{l}
\text { Input } \\
\text { Dimension }
\end{array} $& $\begin{array}{l}
\text { Output } \\
\text { Dimension }
\end{array} $& $\begin{array}{l}
\text { Hidden } \\
\text { State }
\end{array}$ & \text { Input } & \text { \# Neurons } \\
\hline 6 & \text{RNN-LSTM} & {[10,r]} & {[10,p]} & \text{None} & \text{Latent Dimension} & p  \\
7 & \text{RNN-LSTM} & {[10,p]} & {[10,p]} & \text{None} & \text{Layer 6 output} & p  \\
8 & \text{RNN-LSTM} & {[10,p]} & {[10,p]} & \text{Layer 6 internal state} & \text{Layer 7 output} & p  \\
9 & \text{RNN-LSTM} & {[10,p]} & {[10,p]} & \text{Layer 7 internal state} & \text{Layer 8 output} & p  \\
10 & \text{Attention} & {[10,p],[10,p]} & {[10,2p]} & - & \text{Layer 7\&9 output} & p  \\
11 & \text{Dense} & {[10,2p]} & {[10,r]} & - & \text{Layer 10 output} & r  \\
\end{tabular}

  \label{tab:LC_prop}
\end{ruledtabular}
\end{table*}

\begin{table*}
\caption{Attributes of transpose convolutional and dense layers in the decoder $\boldsymbol{\Psi}_{D}(.;\theta_D)$.}
\centering
\begin{ruledtabular}
\begin{tabular}{llllllll}
\text { Layer } & $\begin{array}{l}
\text { Layer } \\
\text { Type }
\end{array}$& $\begin{array}{l}
\text { Input } \\
\text { Dimension }
\end{array} $& $\begin{array}{l}
\text { Output } \\
\text { Dimension }
\end{array} $& $\begin{array}{l}
\text { Kernel } \\
\text { Size }
\end{array} $& $\begin{array}{l}
\text { \# filters/ } \\
\text { \# neurons }
\end{array}$ & \text { Stride }\\
\hline 11 & \text{Dense} & {[10,r]} & {[10,64]} & - & 64 & - \\
12 & \text{Dense} &{[10,64]} & {[10, 128]} & - & 128 & - \\
13 & \text{Dense} &{[10,128]} & {[10,512]} & -  & 512 & -\\
 & \text{Reshape} &{[10,512]} & {[10,16,32]} & -  & - & -\\
 & \text{UpSampling 1D} &{[10,16,32]} & {[10,32,32]} & -  & - & -\\
14 & \text{Conv 1D Transpose} & {[10,32,32]} & {[10,64,64]} & [5]& 64&2 \\
 & \text{UpSampling 1D} &{[10,64,64]} & {[10,128,64]} & -  & - & -\\
15 & \text{Conv 1D Transpose} & {[10,128,64]} & {[10,256,1]} & [5]& 1&2 \\
\end{tabular}

  \label{tab:LC_Dec}
\end{ruledtabular}  
\end{table*}

\begin{figure*}
\centering
\includegraphics[width=\textwidth]{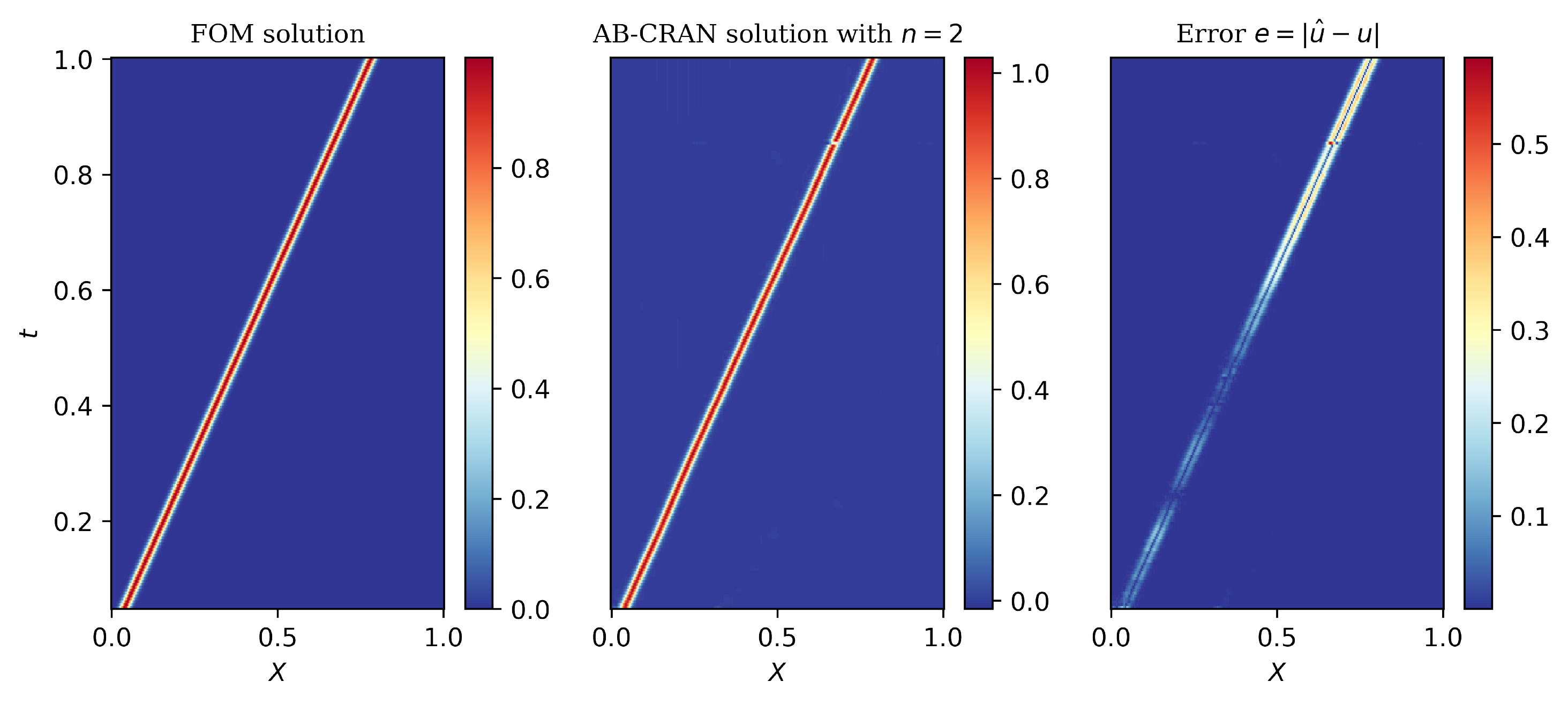}
 \caption{Linear convection problem: Exact solution (left), AB-CRAN solution with n = 2 (center) and error $ e = |\hat{u} - u|$ (right) for the testing parameter $\mu_{test}$ = 0.7875 in the space-time domain.}
\label{fig:char_ABCRAN_mu_0}
\end{figure*}

The first ten time steps of data from the linear convection equation with $\mu = 0.7875$ are fed into the AB-CRAN architecture, and the dimension of the nonlinear trial manifold is set to {$r=2 $ for this case}. Figure~ \ref{fig:char_ABCRAN_mu_0} illustrates both the exact solution and the AB-CRAN approximation for this instance of the testing parameter.
The denoising-based AB-CRAN solution with {$r=2$} accurately captures the amplitude and predicts the wave velocity. 

\subsubsection{Impact of AB-CRAN on time series prediction}
\begin{figure*}
\centering
\includegraphics[width=\textwidth]{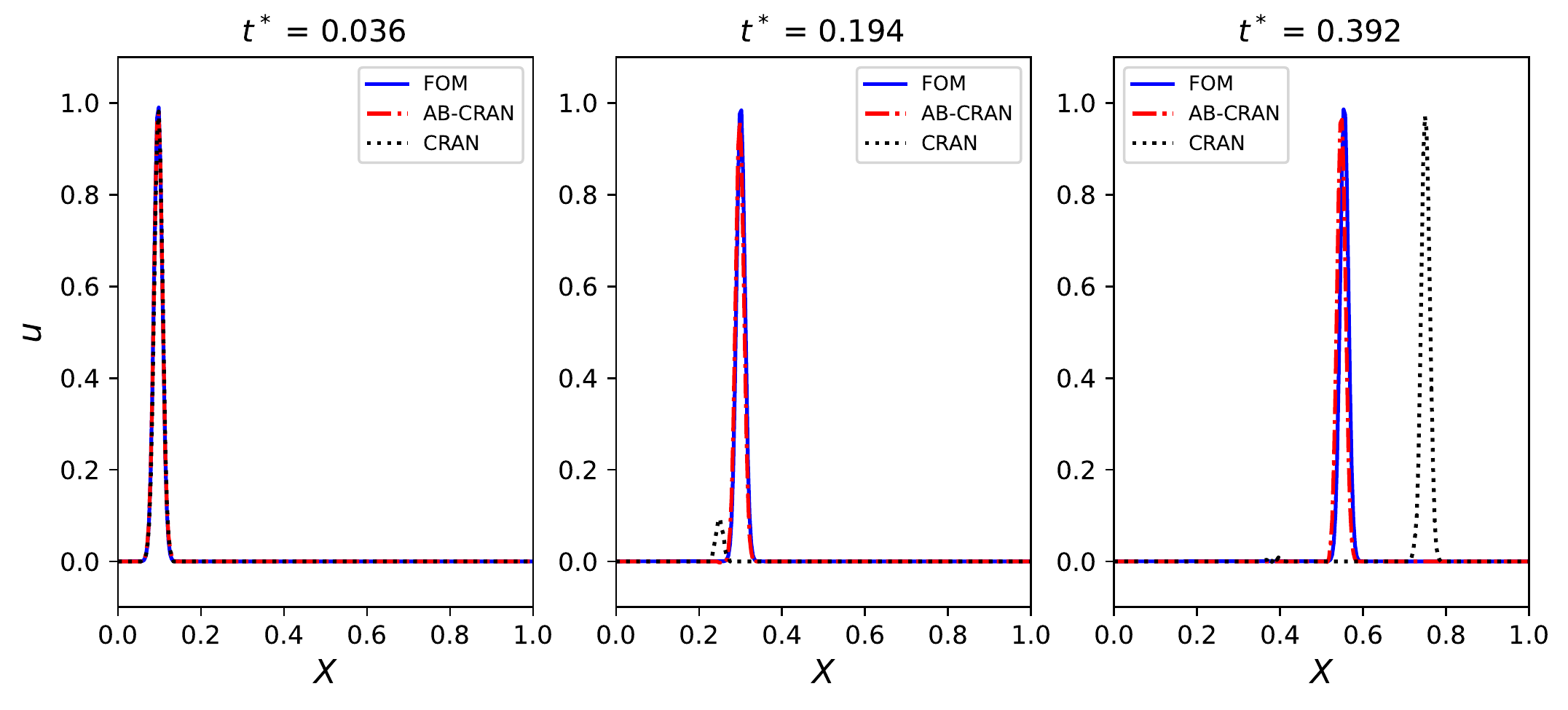}
 \caption{Linear convection problem: Comparison of full-order model solution, CRAN, and AB-CRAN solution at three time instants ($t*=[0.036, 0.194, 0.392]$), where non-dimensional time $t^*= t \mu/L$. 
}
\label{fig:FOM_ABCRAN_CRAN_MU0}
\end{figure*}

\begin{figure*}
\centering
\includegraphics[width=0.7\textwidth]{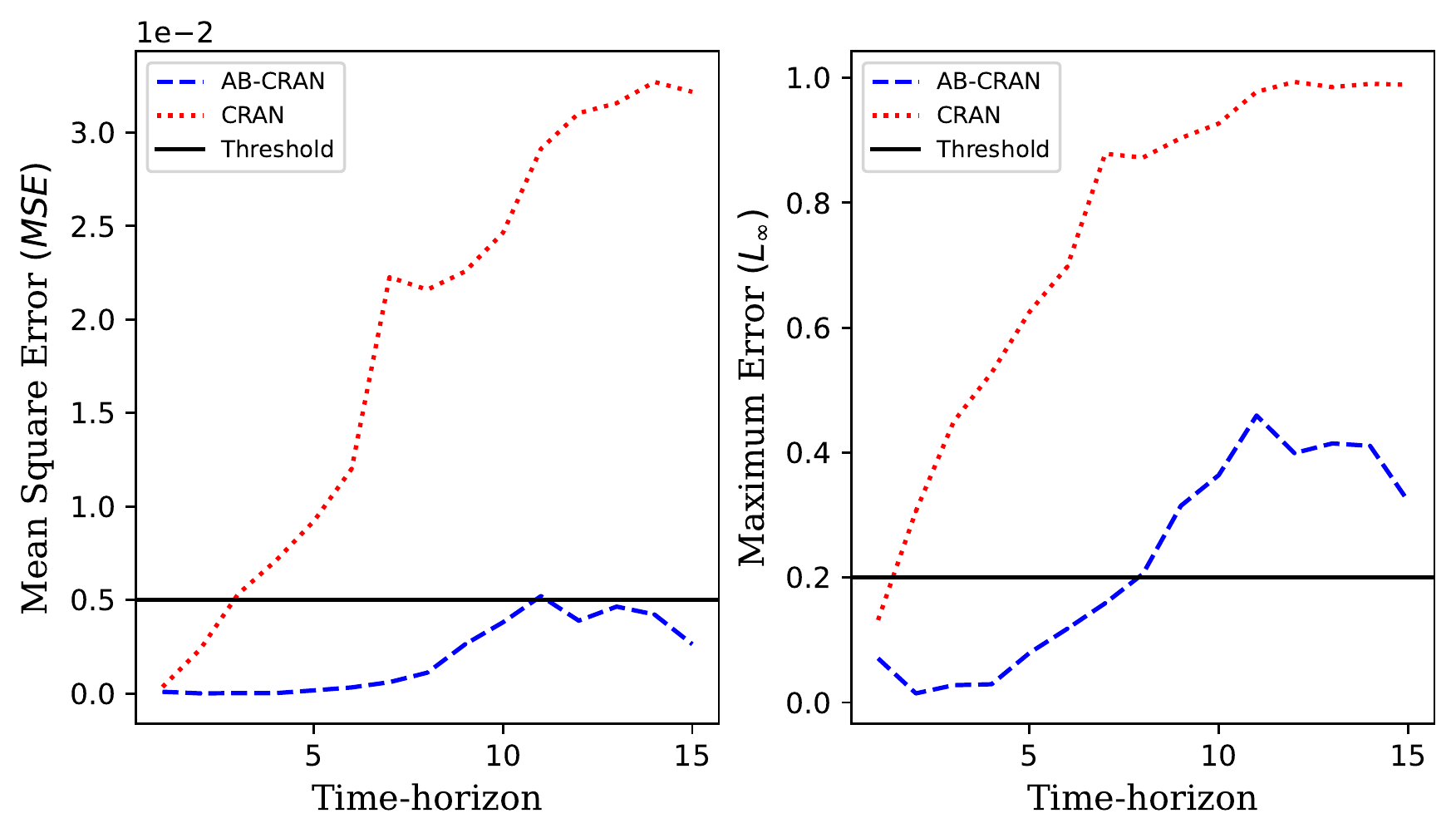}
 \caption{Linear convection problem: Comparison of mean squared error and maximum error of AB-CRAN with CRAN
}
\label{fig:error_MU0}
\end{figure*}

We first examine the AB-CRAN's time series prediction capability to that of the CRAN network. 
We consider a value of 0.7875 for the wave phase speed and aim to accurately predict wave propagation for this parameter case.
We consider the mean squared error, and the maximum error ($L_\infty$) criterion to assess the accuracy of the predictions. While the mean squared error is given by:
\begin{equation}
\operatorname{MSE}(\mathbf{u}, \hat{\mathbf{u}})=\sum_{i=1}^{N}\frac{(\hat{\mathbf{u}}_i^j-\mathbf{u}_i^j)^{2}}{N},
\label{eqn:mse}
\end{equation}
 the error $L_\infty$ is:
\begin{equation}
\operatorname{L_\infty}(\mathbf{u}, \hat{\mathbf{u}})=\operatorname{max}(|\hat{\mathbf{u}}_i^j-\mathbf{u}_i^j|),
\label{eqn:max}
\end{equation}
where $N$ is the spatial degrees of freedom.
Figure \ref{fig:FOM_ABCRAN_CRAN_MU0} illustrates the predictions for non-dimensional time ($t^*=\frac{t\mu}{L}$) values of 0.036, 0.194, and 0.392.
We predict ten-time steps using a sequence-to-sequence mapping model and refer to a sequence of ten-time steps as one-time horizon. Thus, the first, fifth, and tenth time horizons correspond to the tenth,  fiftieth, and hundredth time steps forward, respectively. The results in Fig. \ref{fig:FOM_ABCRAN_CRAN_MU0} demonstrate that the AB-CRAN precisely captures the peak amplitude and wave speed for the testing time steps. 
On the other hand, CRAN architecture with plain long short-term memory cell evolver struggles to capture wave propagation phenomenon beyond first time horizon.
In Fig. \ref{fig:error_MU0}, the mean squared error  and $L_\infty$ error norm of the CRAN and AB-CRAN predictions are compared. 
In comparison to the CRAN, the AB-CRAN models have a lower mean squared error and $L_\infty$.
The results indicate that the AB-CRAN network can significantly reduce the error of CRAN predictions for the linear convection equation.
Mean squared error is negligible in AB-CRAN prediction when compared to the CRAN procedure over the entire time period of prediction. The mean squared error and the maximum error from the CRAN procedure increase with the time horizon, indicating that the error accumulates, whereas the error from our proposed AB-CRAN remains negligible and less than the threshold value. 
{This confirms that the AB-CRAN procedure learns the linear convection equation more effectively than the CRAN and the implicit biases of the convolutional autoencoder and the attention-based time integration are necessary for learning wave propagation}.

\subsubsection{Effect of denoising-based autoencoder}
\begin{figure*}
\centering
\includegraphics[width=0.8\textwidth]{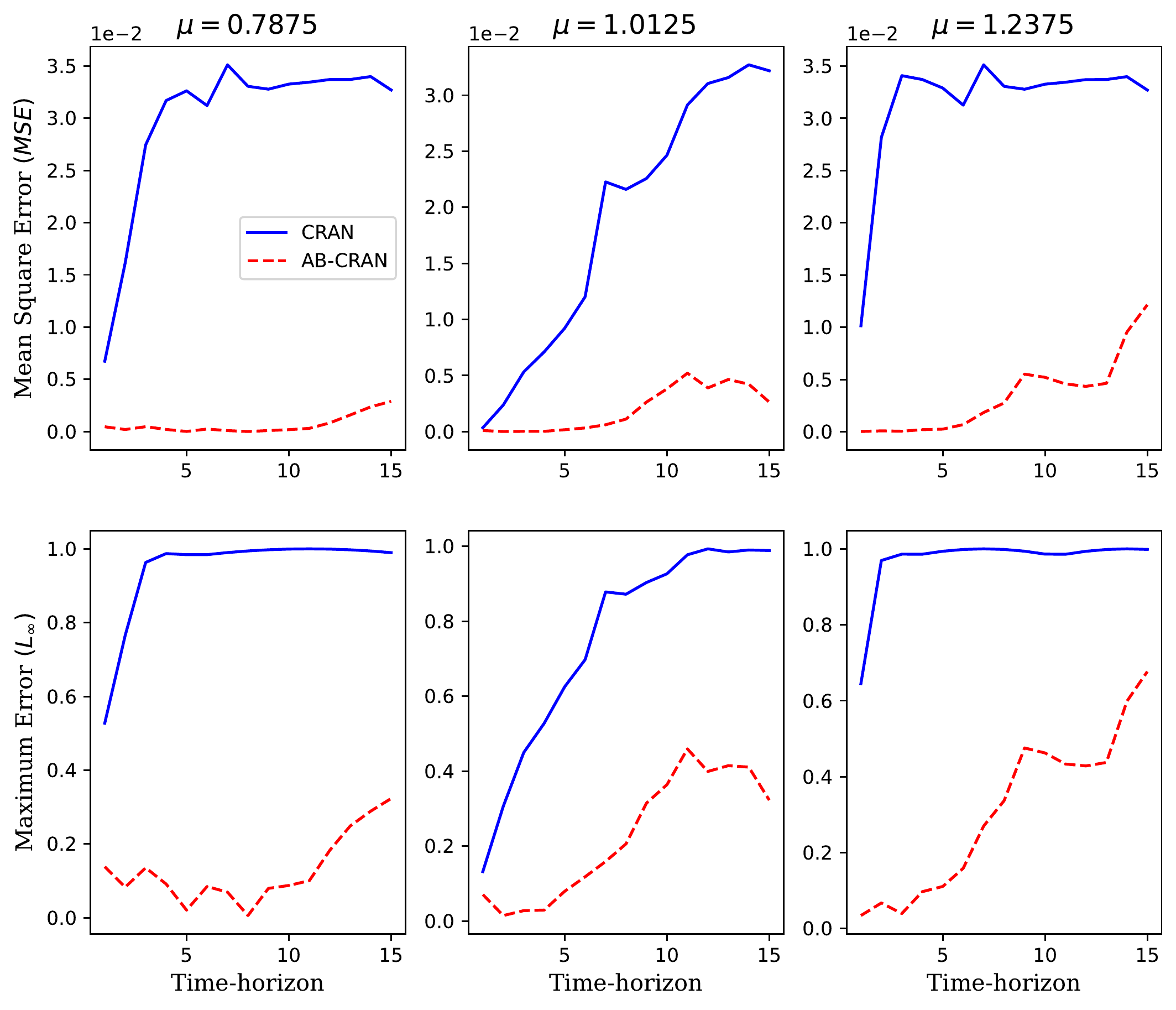}
 \caption{Linear convection problem: Comparison of generalisation error across the parameter space $\mu \in [0.7875, 1.2375]$. }
\label{fig:LC_denoising_generalisation_error}
\end{figure*}

In this section, we will examine the effect of a denoising autoencoder in the AB-CRAN on the generalization of predictions for different parameter values.
One of the main advantages of using a denoising-based autoencoder is the inherent regularization it provides, which prevents over-fitting and improves generalization on the test dataset.
To study the effects of the denoising autoencoder, predictions for three test cases were obtained from the network with denoising training. 
Test Case 1 considers a case where $\mu = 0.7875$, wave phase speed is less than one, Test Case 2 considers a case where $\mu = 1.0125$, wave phase speed is close to one, and Test Case 3 considers a case where $\mu = 1.2375$, wave phase speed is greater than one.
Figure (\ref{fig:LC_denoising_generalisation_error}) shows that the mean squared error and $L_\infty$ error norm for all three test cases are less for AB-CRAN compared to CRAN which shows that the AB-CRAN generalizes better for different parameter instances. The denoising autoencoder as expected provides better results on the entire parameter space and is a more general model.

In addition to the mean squared error (Eq. (\ref{eqn:mse})) and the maximum error (Eq. (\ref{eqn:max})), we consider the time average value of the mean squared error and the maximum errors as an alternative metric to assess the accuracy of the predictions for different parameters, which are given by
\begin{equation}
 <MSE(\mathbf{u},\hat{\mathbf{u}})>=\sum_{j=1}^{N_t}\left(\sum_{i=1}^{N}\frac{(\hat{\mathbf{u}}_i^j-\mathbf{u}_i^j)^{2}}{N}\right)/({N_t}),
\label{eqn:TVmse}
\end{equation}
and
\begin{equation}
<{L_\infty}(\mathbf{u}, \hat{\mathbf{u}})>=\sum_{j=1}^{N_t}\frac{\operatorname{max}(|\hat{\mathbf{u}}_i^j-\mathbf{u}_i^j|)}{N_t},
\label{eqn:TVmax}
\end{equation}
where $N$ is the spatial degrees of freedom, and $N_t$ is number of time steps, and $< >$ denotes the time averaging.
In a nutshell, it can be seen in  Table \ref{table:mse_Linf_LC} that the AB-CRAN reduces the mean squared error by an order of magnitude in comparison to CRAN while it reduces the maximum error by four times.
{As a function of the data size, a comparison of time averaged mean squared error for training, validation, and test data-set is summarized in Table \ref{table:amt_data_LC}.}
\begin{table}[h]
\centering
\caption{Linear convection problem: Comparison of the time averaged mean squared error and the maximum error between AB-CRAN and CRAN for various parameter cases.}
\begin{tabular}{ccccc}
\toprule
Parameter & \multicolumn{2}{c}{$<MSE>$} & \multicolumn{2}{c}{$<L_\infty>$}\\
        $\mu$ & CRAN & AB-CRAN & CRAN  & AB-CRAN \\
        \midrule
$0.7875$  & 0.0254 & 0.0012 & 0.9009 & 0.1776\\
$1.0125$ & 0.0254 & 0.0017 & 0.8926 & 0.2215\\
$1.2375$ & 0.0229 & 0.0034 & 0.7203 & 0.2692\\
\bottomrule
\end{tabular}

\label{table:mse_Linf_LC}
\end{table}

\begin{table}[h]
\centering
\caption{{Linear convection problem: Time averaged mean squared error for training, validation and test set as function of amount of data.}}
\begin{tabular}{ccc}
\toprule
Data-set & {Amount of data} & {$<MSE>$}\\
\midrule
Training  & 2,896 seqs (148MB) & $1.28\times10^{-4}$ \\
Validation & 724 seqs (37MB) & $2.03\times10^{-4}$ \\
Test & 361 seqs (15MB) & $2.37\times10^{-4}$\\
\bottomrule
\end{tabular}
\label{table:amt_data_LC}
\end{table}
\subsection{Viscous Burgers equation}
In this section, we consider the viscous Burgers' equation as a model for nonlinear wave propagation. 
Consider the following parametric partial differential equation: 
\begin{equation}
\frac{\partial{u}}{\partial{t}}+u \frac{\partial u}{\partial x}=\nu \frac{\partial^{2} u}{\partial x^{2}},
\end{equation}
with Dirichlet boundary conditions and following initial condition.
\begin{align}
    u(x, 0)&=\frac{x}{1+\sqrt{\frac{1}{t_{0}}} \exp \left(\operatorname{Re} \frac{x^{2}}{4}\right)} \quad \text{on}\quad [0, L], \\ u(0, t)&=u(L, t)=0,
\end{align}
We define $Re = \frac{1}{\nu}$, which varies from 1000 to 4000, and consider $N_{Re} = 7$ training parameter instances distributed uniformly over this range. 
We set the physical domain length $L \in [0,1]$ and discretize it into $256$ spatial points. 
We set $t_{max} = 2$ and discretize it into 200 time steps. The analytical solution for the viscous Burgers' equation with the specified initial condition is given  by
\begin{equation}
    u(x, t)=\frac{\frac{x}{t+1}}{1+\sqrt{\frac{t+1}{t_{0}}} \exp \left(\operatorname{Re} \frac{x^{2}}{4 t+4}\right)},
\end{equation}
where $t_{0}=\exp (Re / 8)$.
Because of convection-dominated behavior, the viscous Burgers equation can produce discontinuous solutions.  

\begin{figure*}
\centering
\includegraphics[width=\textwidth]{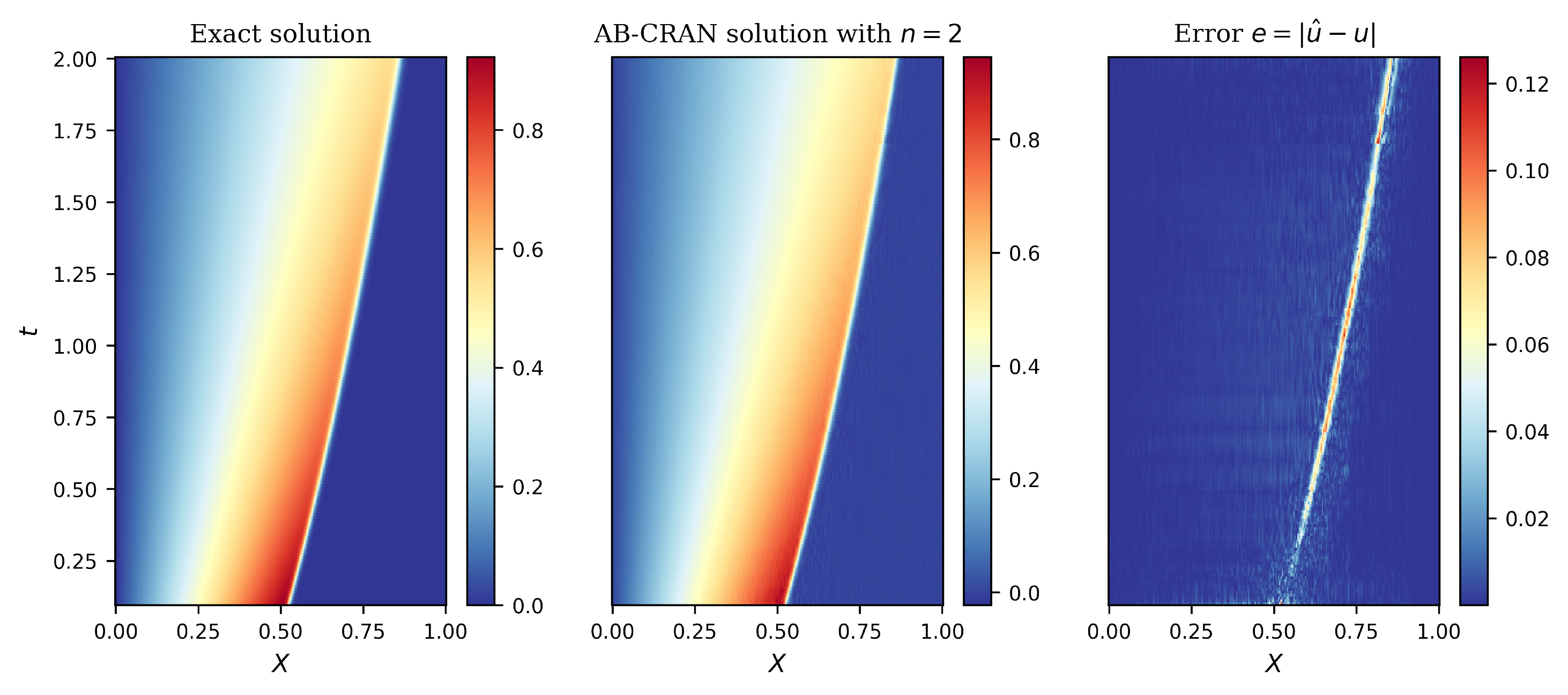}
 \caption{Nonlinear viscous Burgers problem: Exact solution (left), AB-CRAN solution with n = 2 (center) and error $ e = |\hat{u} - u|$ (right) for
the testing-parameter instance $Re$ = 1100 in the space-time domain}
\label{fig:Char_ABCRAN_VB_Re_1100}
\end{figure*}

In this test case of nonlinear convection, the same neural network architecture with  15-layers of AB-CRAN is used. We also set the dimension of the reduced manifold to two {($r=2$)}.
As input, the first ten time steps of the viscous Burgers' equation with $Re = 1100$ are used.  Figure \ref{fig:Char_ABCRAN_VB_Re_1100} shows both the exact solution and the AB-CRAN approximation for this particular instance of the testing parameter. Denoising AB-CRAN solution with {$r=2$} accurately captures the nonlinear wave propagation. {A comparison of time averaged mean squared error for training, validation, and test data-set as a function of amount of data is given in Table \ref{table:amt_data_VB}.}
\begin{table}[h]
\centering
\caption{{Viscous Burgers problem: Time averaged mean squared error for training, validation and test set as function of amount of data.}}
\begin{tabular}{ccc}
\toprule
Data-set & {Amount of data} & {$<MSE>$}\\
\midrule
Training  & 2,751 seqs (141MB) & $2.2\times10^{-5}$ \\
Validation & 688 seqs (35MB) & $5.8\times10^{-5}$ \\
Test & 342 seqs (13MB) & $6.2\times10^{-5}$\\
\bottomrule
\end{tabular}
\label{table:amt_data_VB}
\end{table}

\subsubsection{AB-CRAN predictions for varying Reynolds number}
\begin{figure*}
  \centering
  \includegraphics[width=.7\linewidth]{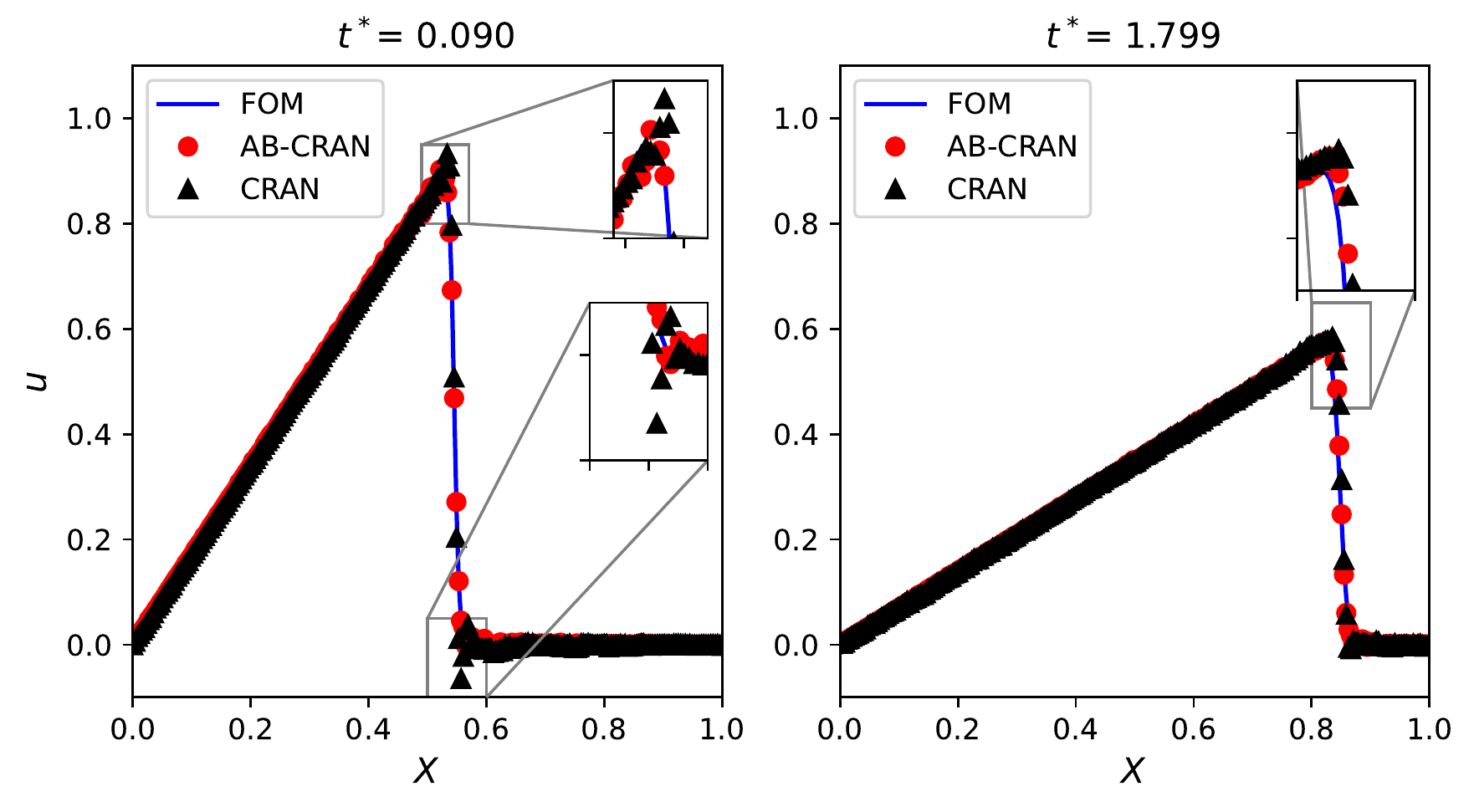}  
\\ (a) \\
  \includegraphics[width=.7\linewidth]{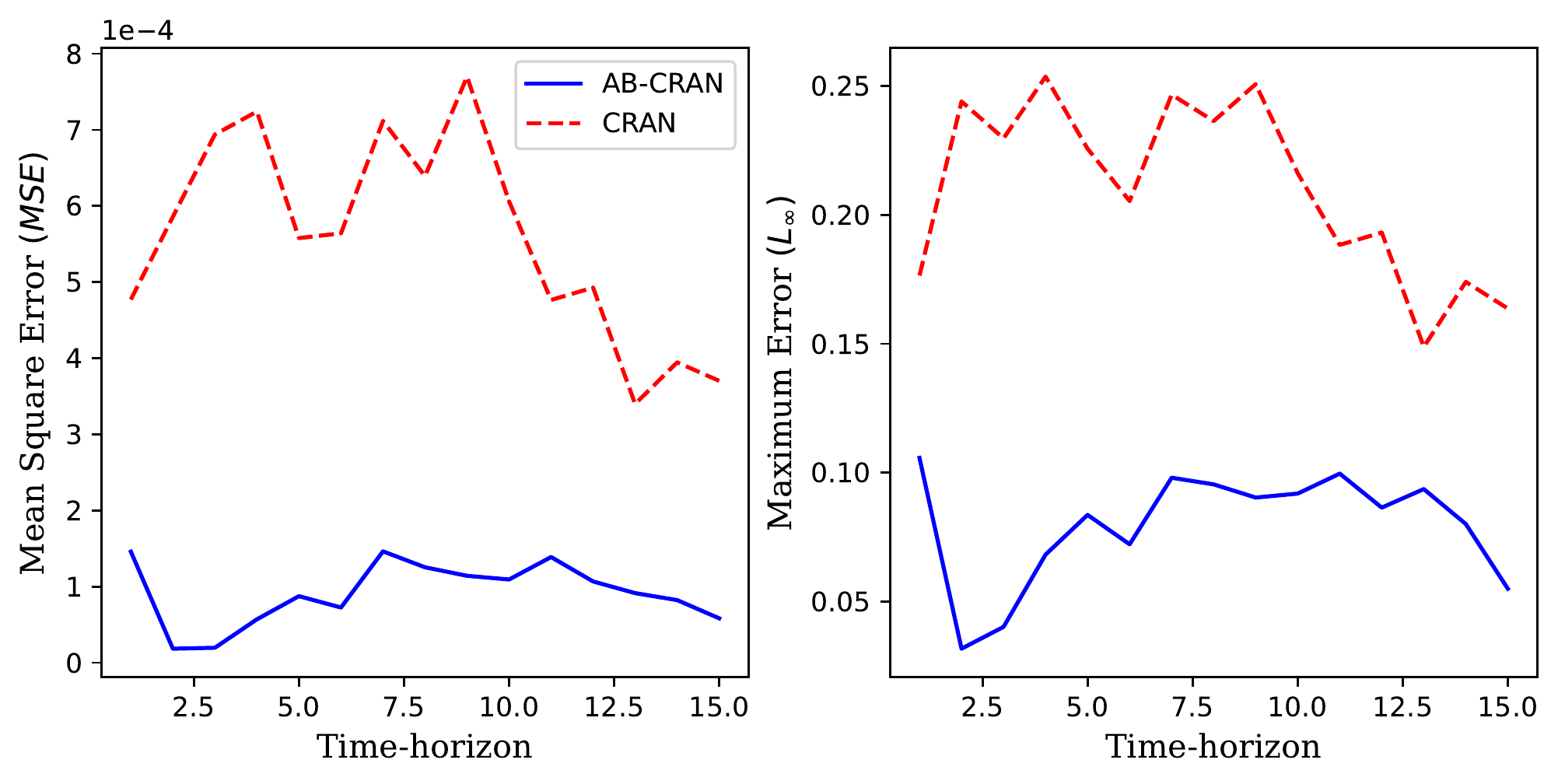}  
\\  (b)
\caption{Error plots and predictions from AB-CRAN and CRAN for Re = 1100. (a) Shows AB-CRAN and CRAN predictions for Re = 1100 at time steps $t^*=\{0.09,1.79\}$. (b) Illustrates mean squared error and maximum error from AB-CRAN and CRAN for Re = 1100.}
\label{fig:fig_Visc_Burgers_Re_1100}
\end{figure*}

In this section, we will assess the predictions of our AB-CRAN framework for varying Reynolds numbers between 1000 and 4000. In particular, we will consider two test cases namely: Test Case 1 corresponding to a Reynolds number of 1100 and Test Case 2 with a Reynolds number of 3600.
Figure~\ref{fig:fig_Visc_Burgers_Re_1100} illustrates the predicted value and the error for Test Case 1 with Reynolds number 1100. As it can be seen that the AB-CRAN accurately captures both nonlinear wave propagation and the discontinuity feature in the solution. While the CRAN network exhibits oscillation near the discontinuity, our AB-CRAN technique is more effective at learning the physics of the viscous Burgers' equation.

The mean squared error and the maximum error ($L_{\infty}$) of the AB-CRAN and CRAN predictions are compared in Fig. \ref{fig:fig_Visc_Burgers_Re_1100}.  
The AB-CRAN provides a relatively smaller error compared to the CRAN architecture. The results show that the  AB-CRAN reduces the error of CRAN predictions by approximately 50\%  for the testing parameter of $Re_{1100}$.

\begin{figure*}
  \centering

  \includegraphics[width=.7\linewidth]{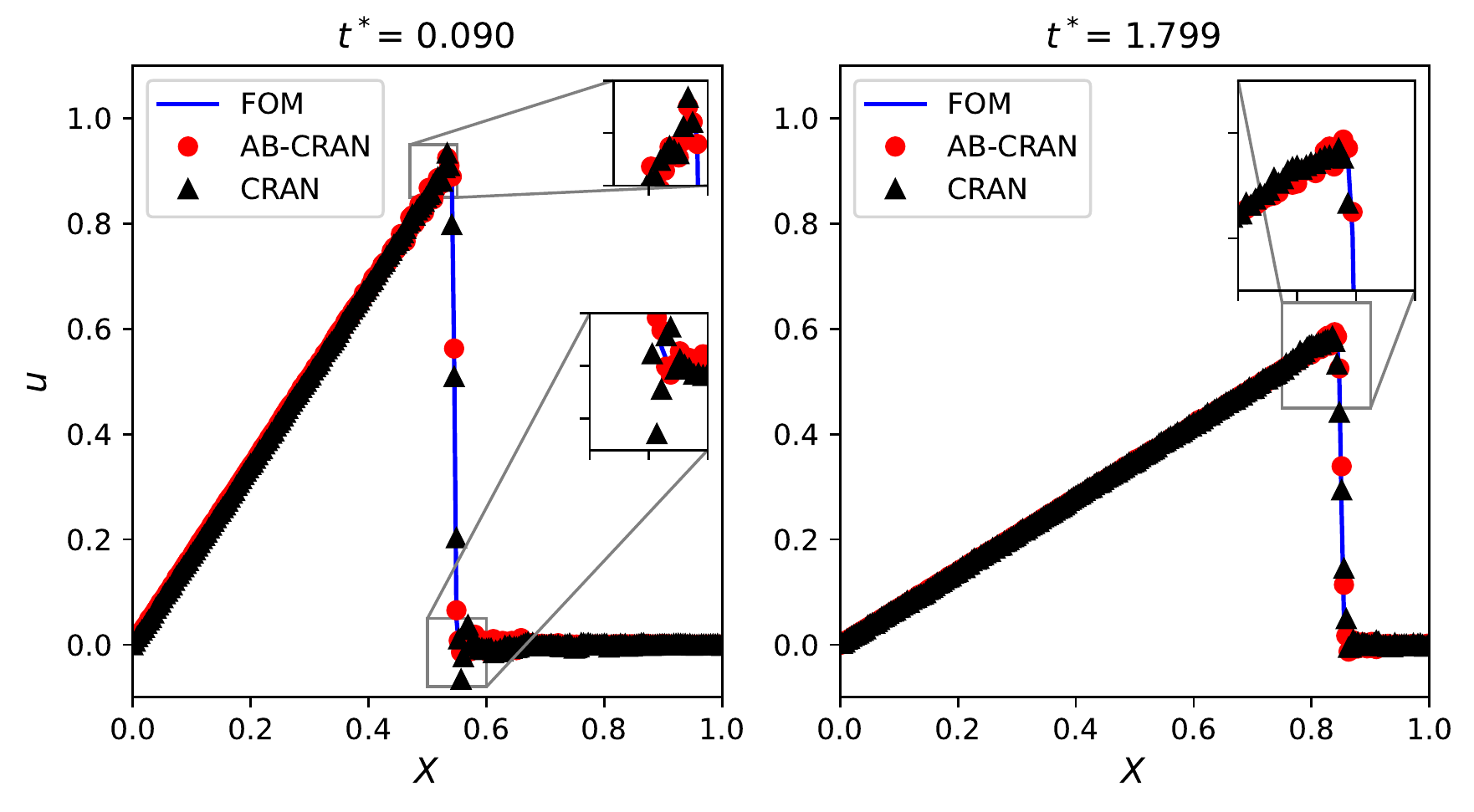}  
\\ (a) \\
  \includegraphics[width=.7\linewidth]{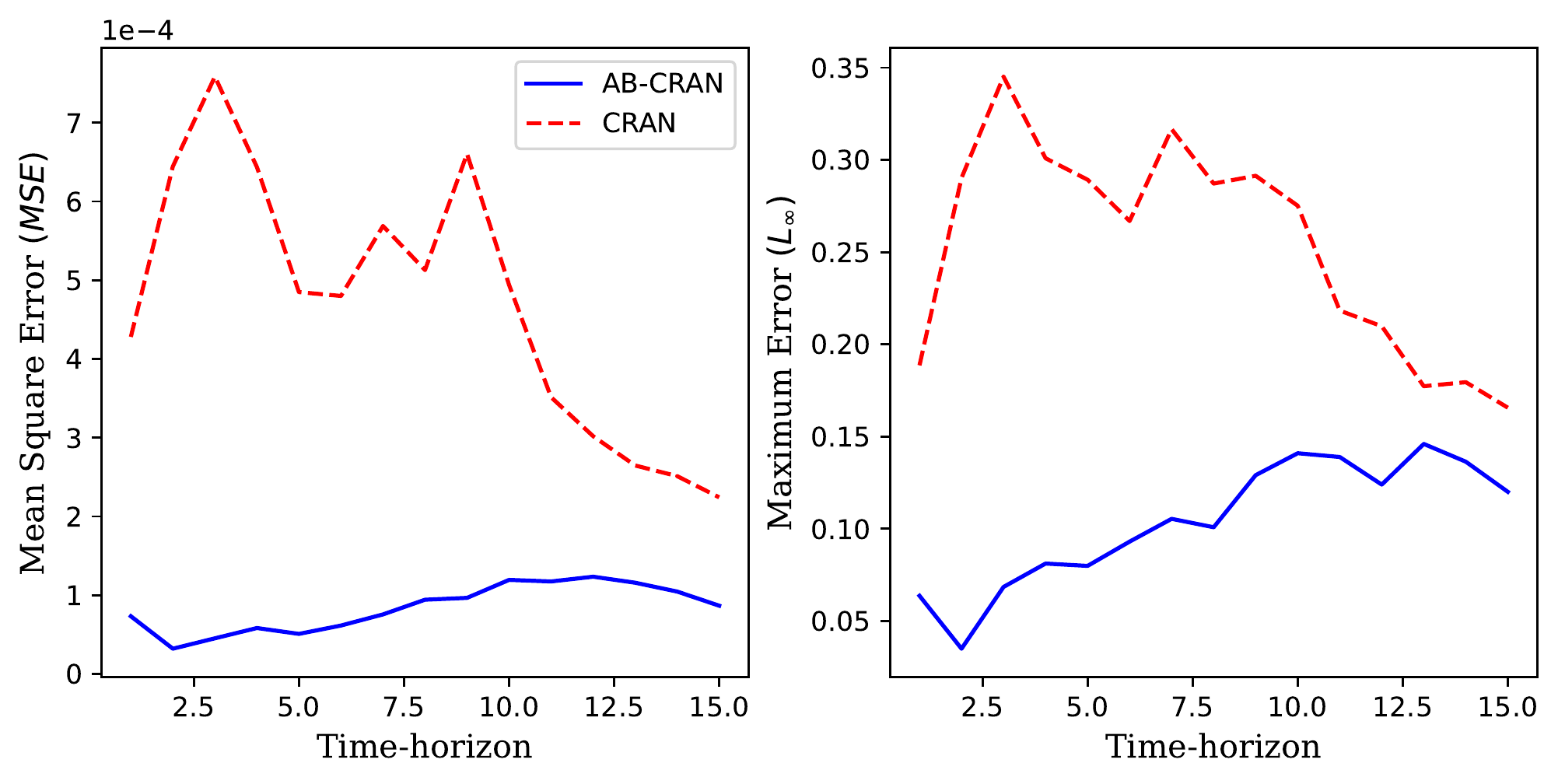}  
\\ (b)

\caption{Nonlinear viscous Burger problem: Error plots and predictions from AB-CRAN and CRAN at Re = 3600. (a) Shows AB-CRAN and CRAN predictions for Re = 3600 at time steps $t^*=\{0.09,1.79\}$. (b) Illustrates MSE and maximum error from AB-CRAN and CRAN for Re = 3600.}
\label{fig:fig_Visc_Burgers_Re_3600}
\end{figure*}

The predicted value and errors for Test Case 2, with Reynolds number 3600 exhibit the same trend, similar to Test Case 1. The AB-CRAN accurately models nonlinear wave propagation and discontinuity, whereas the CRAN network exhibits oscillation near the discontinuity.
Moreover, for the testing parameter $Re_{3600}$, fig. \ref{fig:fig_Visc_Burgers_Re_3600} shows that the AB-CRAN reduces the error of CRAN predictions approximately by 50\%.
The errors summarized in table \ref{table:mse_Linf_VB} show that AB-CRAN reduces the mean squared error by an order of magnitude in comparison to CRAN, and it reduces the maximum error by three times.

\begin{table}[H]
\centering
\caption{Nonlinear viscous Burgers problem: Comparison of time averaged mean squared error and maximum error between AB-CRAN and CRAN for various parameter cases.}
\begin{tabular}{ccccc}
\toprule
{Parameter} & \multicolumn{2}{c}{$<MSE>$} & \multicolumn{2}{c}{$<L_\infty>$}\\
       $Re$  & CRAN & AB-CRAN & CRAN  & AB-CRAN \\
        \midrule
$1100$  & $4.873\times10^{-4}$ & $8.118\times10^{-5}$ & 0.192 & 0.073 \\
$3600$ & $3.068\times10^{-4}$  & $7.433\times10^{-5}$ & 0.202 & 0.096\\
\bottomrule
\end{tabular}
\label{table:mse_Linf_VB}
\end{table}
\subsection{2D Shallow Water Wave Propagation}
We now consider the 2D shallow water model given by Saint-Venant equations. 
The PDE system provides a hydrodynamic model that calculates the flow velocity and the water level over a two-dimensional domain. 
It takes into account the various forces influencing and accelerating the flow. 
The 2D horizontal Saint-Venant mathematical model arises from the vertical integration of the 3D Navier-Stokes equations with various assumptions such as the vertical pressure gradient is nearly hydro-static (i.e., long waves) and the horizontal length scale is much larger than the vertical length scale. 

\begin{figure*}
\centering
\includegraphics[width=0.8\textwidth]{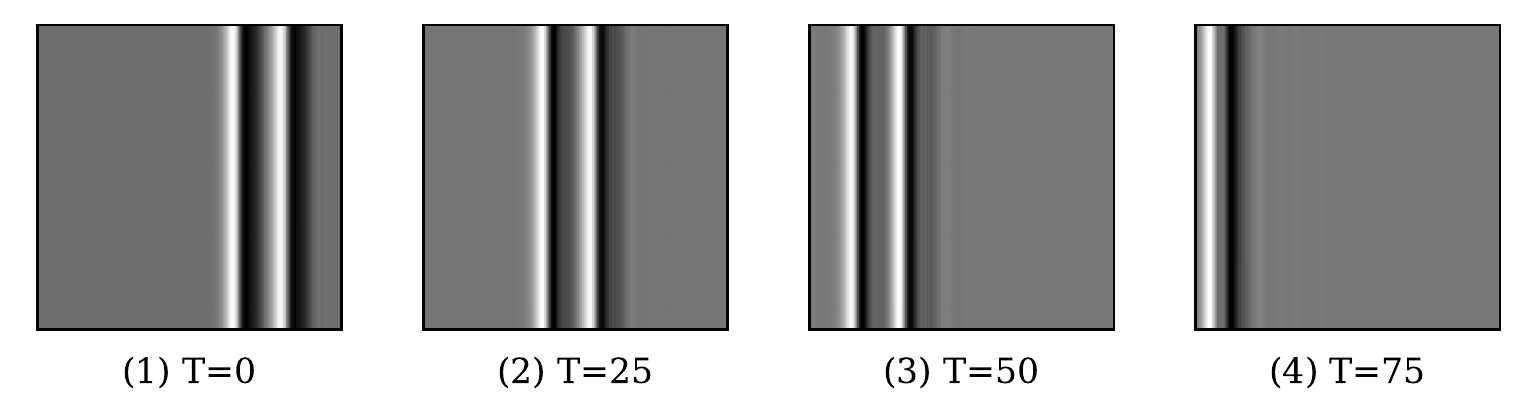}
 \caption{2D Saint-Venant problem: Illustration of the propagation of a plane wave. $T$ represents the number of time steps from the initial condition.}
\label{fig:FOM_PW}
\end{figure*}

The Saint-Venant model comprises the equation for mass conservation and the two equations of momentum conservation and can be written in a non-conservative form as:
\begin{equation}
\left.
\begin{aligned}
\frac{\partial h}{\partial t}+\pder{x}{\left((H+h) u\right)}+\pder{y}{((H+h) v)} &=0, \\
\frac{\partial u}{\partial t}+u \frac{\partial u}{\partial x}+v \frac{\partial u}{\partial y}+g \frac{\partial h}{\partial x}-\nu\left(\frac{\partial^2 u}{{\partial x}^2}+\frac{\partial^2 u}{{\partial y}^2}\right) &=0, \\
\frac{\partial v}{\partial t}+u \frac{\partial v}{\partial x}+v \frac{\partial v}{\partial y}+g \frac{\partial h}{\partial y}-\nu\left(\frac{\partial^2 v}{{\partial x}^2}+\frac{\partial^2 v}{{\partial y}^2}\right) &=0, 
\end{aligned}
\right\}
\end{equation}
here, $u$ and $v$ are the velocities in the $x$ and $y$ direction, respectively, $H$ denotes the reference water height and $h$ is the deviation from this reference, $g$ is the acceleration due to gravity, $\nu$ is the kinematic viscosity. 
The solid wall boundary conditions are used along its perimeter.
We use a plane wave as the initial condition. 
The data were generated using Python package TriFlow \cite{nicolas_cellier_2017_584101}. 
An example of wave pattern evolution generated by the numerical solver is illustrated in Fig. \ref{fig:FOM_PW}.
The dataset was generated by varying the initial location of the plane wave. 
We select $t_{max} = 1$ and discretize it into 100 time steps.
The images were rendered with $184\times184$ pixels.

\subsubsection{Data-driven predictions via AB-CRAN}
%
The architecture of the neural network used for this test case is similar to the first two cases. 
{In particular, we augment the network architecture to handle the two-dimensional input data and the reduced dimension is set to eight ($r=8$).} 
We choose a 15-layers  convolutional recurrent autoencoder.  The architecture of the network is identical to the one-dimensional case, only two-dimensional convolution and max-pooling are used instead of one-dimensional operations. 
The total number of trainable parameters (i.e., weights and biases) of the neural network is 4,504,785.
{The model was trained from scratch with TensorFlow \cite{tensorflow2015-whitepaper} using single NVIDIA P100 Pascal graphical processing unit, and 16 cores Intel E5-2683 v4 Broadwell central processing unit with 64GB of system's memory. The training converges in approximately 300 epochs and 7 hours of wall clock time.}

\begin{figure*}
\centering
\includegraphics[width=0.7\textwidth]{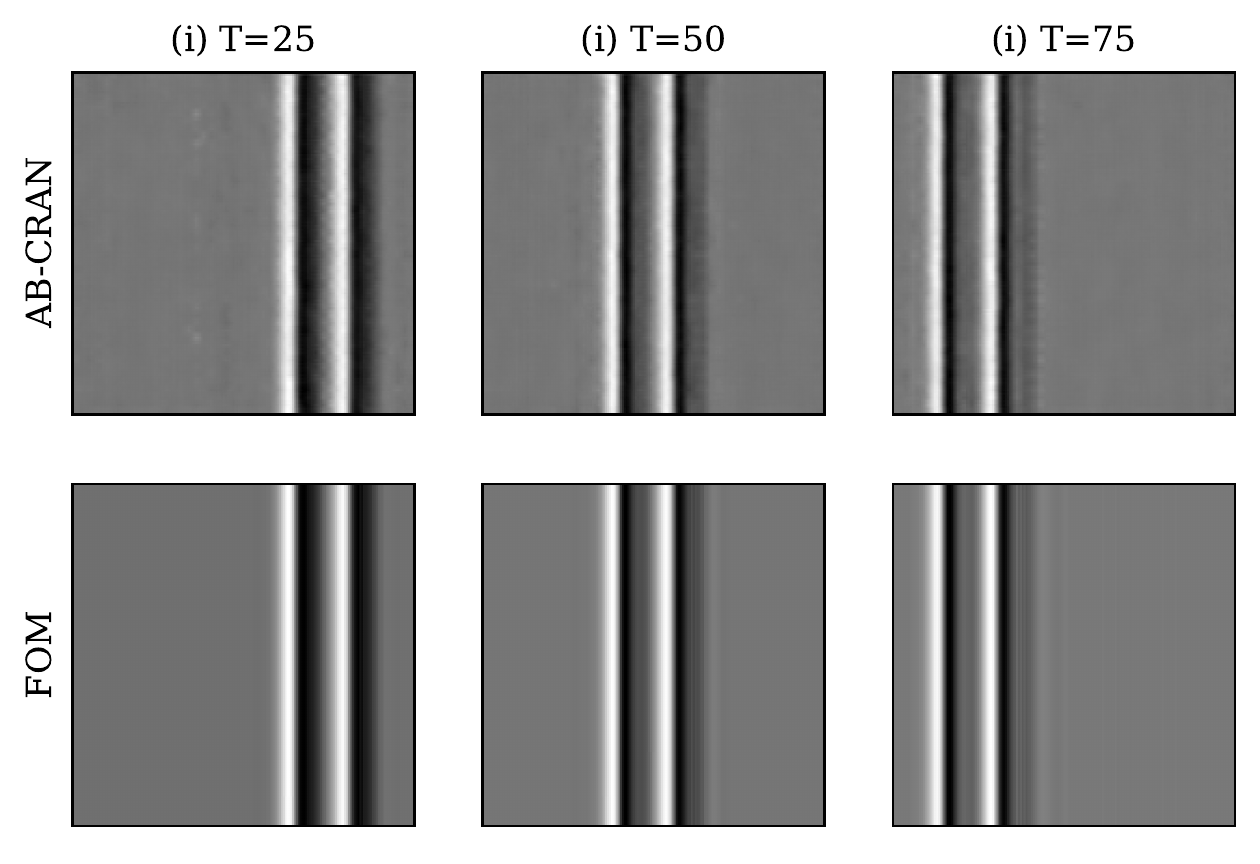}
 \caption{2D Saint-Venant shallow water problem: predicted two-dimensional spatial patterns from AB-CRAN and full-order model solution from a high-fidelity numerical solver.}
\label{fig:pred_2D}
\end{figure*}

\begin{table*}
\caption{Saint-Venant shallow water problem: Network architecture.}
\centering
\begin{ruledtabular}
\begin{tabular}{lllllll}
\textbf{Encoder}\\
\hline \text{Layer} &$\begin{array}{l}
\text{Layer}\\
\text {Type}
\end{array}$&$\begin{array}{l}
\text {Input} \\
\text {Dimension}
\end{array}$ &$\begin{array}{l}
\text {Output} \\
\text {Dimension}
\end{array}$ &$\begin{array}{l}
\text {Kernel} \\
\text {Size}
\end{array}$ &$\begin{array}{l}
\text {\# filters/} \\
\text {\# neurons}
\end{array}$ &\text {Stride}\\
\hline 1 & \text{Conv 2D} & {[10, 184, 184, 1]} & {[10, 92, 92, 64]} & {[5]} & 64 & 2 \\
 & \text{MaxPool 2D} &{[10, 92, 92, 64]} & {[10, 46, 46, 64]} & - & - & - \\
2 & \text{Conv 2D} & {[10, 46, 46, 64]} & {[10, 23, 23, 32]} & {[5]} & 32 & 2 \\
 & \text{Flatten} &{[10, 23, 23, 32]} & {[10, 16928]} & - & - & - \\
3 & \text{Dense} &{[10, 16928]} & {[10, 128]} & -  & 128 & -\\
4 & \text{Dense} &{[10, 128]} & {[10, 64]} & -  & 64 & -\\
5 & \text{Dense} & {[10, 64]} & {[10, r]} & -& r&- \\
\hline
\textbf{Evolver}\\
\hline
\text { Layer } & $\begin{array}{l}
\text { Layer } \\
\text { Type }
\end{array}$& $\begin{array}{l}
\text { Input } \\
\text { Dimension }
\end{array}$ & $\begin{array}{l}
\text { Output } \\
\text { Dimension }
\end{array} $& $\begin{array}{l}
\text { Hidden } \\
\text { State }
\end{array}$ & \text { Input } & \text { \# Neurons } \\
\hline 6 & \text{RNN-LSTM} & {[10,n]} & {[10,p]} & \text{None} & \text{Latent Dimension} & p  \\
7 & \text{RNN-LSTM} & {[10,p]} & {[10,p]} & \text{None} & \text{Layer 6 output} & p  \\
8 & \text{RNN-LSTM} & {[10,p]} & {[10,p]} & \text{Layer 6 internal state} & \text{Layer 7 output} & p  \\
9 & \text{RNN-LSTM} & {[10,p]} & {[10,p]} & \text{Layer 7 internal state} & \text{Layer 8 output} & p  \\
10 & \text{Attention} & {[10,p],[10,p]} & {[10,2p]} & - & \text{Layer 7\&9 output} & p  \\
11 & \text{Dense} & {[10,2p]} & {[10,r]} & - & \text{Layer 10 output} & r  \\
\hline
\textbf{Decoder}\\
\hline
\text { Layer } & $\begin{array}{l}
\text { Layer } \\
\text { Type }
\end{array}$& $\begin{array}{l}
\text { Input } \\
\text { Dimension }
\end{array}$ & $\begin{array}{l}
\text { Output } \\
\text { Dimension }
\end{array} $&$ \begin{array}{l}
\text { Kernel } \\
\text { Size }
\end{array} $& $\begin{array}{l}
\text { \# filters/ } \\
\text { \# neurons }
\end{array} $& \text { Stride }\\
\hline 11 & \text{Dense} & {[10,r]} & {[10,64]} & - & 64 & - \\
12 & \text{Dense} &{[10,64]} & {[10, 128]} & - & 128 & - \\
13 & \text{Dense} &{[10,128]} & {[10,16928]} & -  & 512 & -\\
 & \text{Reshape} &{[10,16928]} & {[10,23,23,32]} & -  & - & -\\
14 & \text{Conv 2D Transpose} &{[10,23,23,32]} & {[10,46,46,64]} & [5]  & 64 & 2\\
 & \text{UpSampling 2D} & {[10,46,46,64]} & {[10,92,92,64]} & -& -&- \\
15 & \text{Conv 2D Transpose} &{[10,92,92,64]} & {[10,184,184,1]} & [5]  & 1 & 2\\
\end{tabular}

  \label{tab:2D_Arch}
\end{ruledtabular}  
\end{table*}
We generate 10 wave solution from the full-order model and used them as the input. 
Figure \ref{fig:pred_2D} shows the solution from full-order model and AB-CRAN.  It can be seen that the AB-CRAN framework can predict the spatial pattern and the wave amplitude for the Saint-Venant equations at a reasonable accuracy.
{For a single 2D shallow water simulation, Triflow takes about 5 minutes to generate 100 time-steps using finite difference formulation, whereas AB-CRAN during inference takes about 20 seconds to generate 100 time-steps.} {A comparison of time averaged mean squared error for training, validation, and test data-set as a function of amount of data is given in Table \ref{table:amt_data_2D}.}

\begin{table}[h]
\centering
\caption{{2D shallow water problem: Time averaged mean squared error for training, validation and test set as a function of amount of data.}}
\begin{tabular}{ccc}
\toprule
Data-set & {Amount of data} & {$<MSE>$}\\
\midrule
Training  & 18,225 seqs (1.8GB) & $8.08\times10^{-4}$ \\
Validation & 4,556 seqs (450MB) & $0.002$ \\
Test & 2531 seqs (254MB) & $0.005$\\
\bottomrule
\end{tabular}
\label{table:amt_data_2D}
\end{table}

\begin{figure*}
\centering
\includegraphics[width=0.7\textwidth]{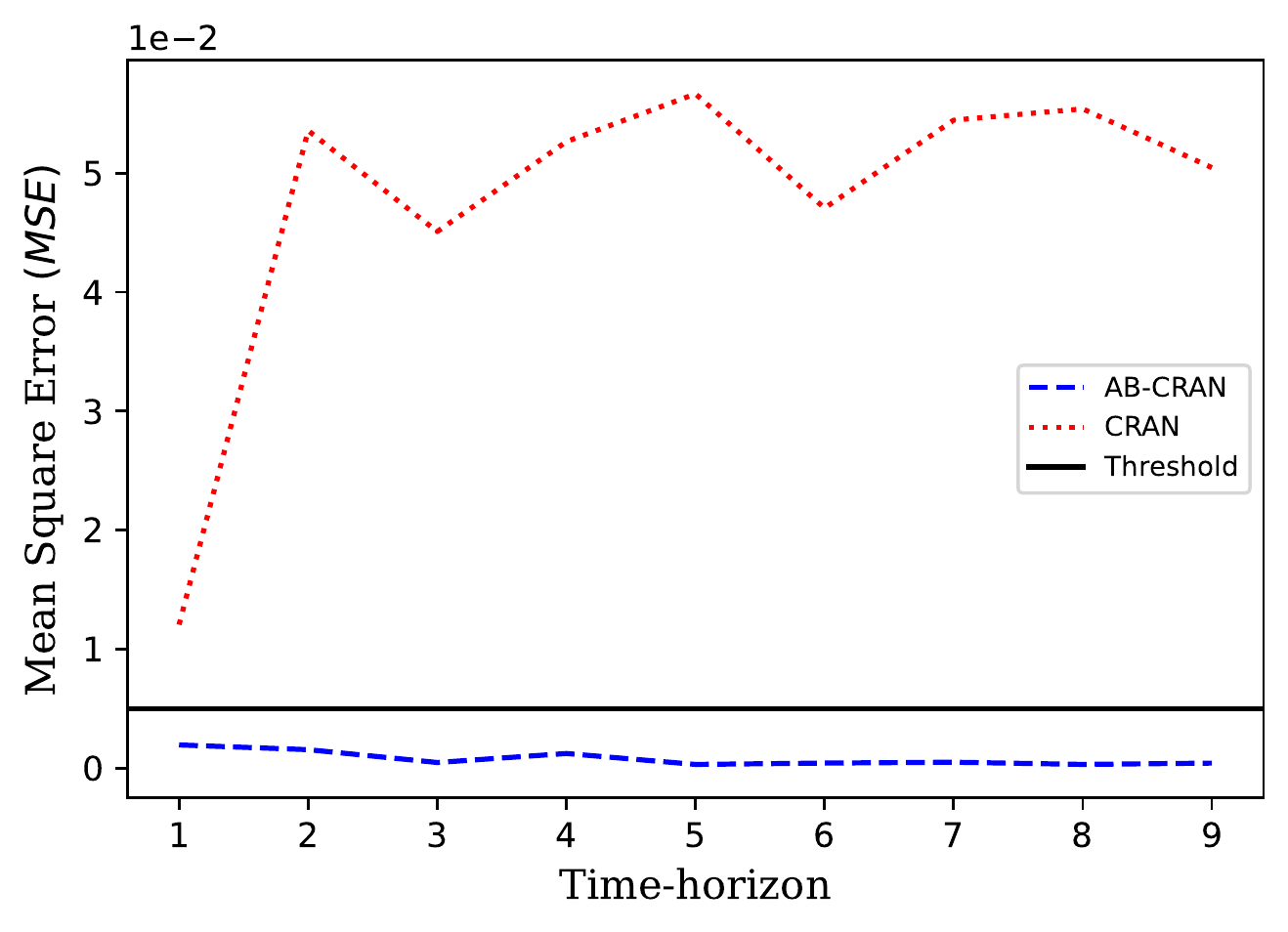}
 \caption{2D Saint-Venant shallow water problem: Comparison of mean square error vs time-horizon for the predicative capability of AB-CRAN and CRAN. While blue curve is the prediction from AB-CRAN, red curve shows the prediction from CRAN. Herein, a sequence of ten time steps is termed as one time-horizon. }
\label{fig:error_2D}
\end{figure*}

In Fig. \ref{fig:error_2D}, the mean squared error of the CRAN and AB-CRAN predictions {with $r=8$} are compared. 
In comparison to the CRAN, the AB-CRAN models exhibit consistently lower mean square error. The results indicate that the AB-CRAN network can significantly reduce the error of CRAN predictions for the two-dimensional cases as well.  
The results suggest that our trained network can perform the wave propagation for the two-dimensional case with minimal hyperparameter tuning hence that the present algorithm confirms the scalability to multi-dimensions. 

\subsection{Discussion}
The solution of a hyperbolic partial differential equation can be integrated in time along the integral curves i.e. characteristics. Information propagates at finite speeds along these low-dimensional characteristic curves and the solution is a superposition of simple wave-like characteristics. Loosely speaking, characteristic curves allow reducing the hyperbolic PDE solution into an ordinary differential equation. The data matrix generated from the hyperbolic PDEs contains essential information about the characteristic curves and the underlying physics of the problem. Passing this data matrix through a deep neural network architecture successively transforms the data matrix and hierarchically extracts the significant spatial and temporal features. 
Particular deep neural network architectures have certain biases to extract a certain type of features, e.g. CNN with max-pooling extracts translational invariant features. 
The stochastic and randomness in the denoising convolutional autoencoder allows the AB-CRAN architecture to further identify the translational invariant low-dimensional manifold similar to projecting the hyperbolic PDE along the characteristic curves. 
Attention-based sequence-to-sequence modeling attempts to learn the trajectory along these curves and transposes convolution projects the solution back to physical space. The physical priors endowed in the AB-CRAN neural network architecture allow it to predict wave propagation for large time horizons.  

The AB-CRAN architecture is capable enough to capture the temporal evolution of the initial disturbance and the different spatial behaviors of the solution across the domain. The attention-based sequence-to-sequence long short-term memory cell for evolving latent dimensions in time captures dependency in long-term sequences and preserves the information for long-time horizons. The numerical integration methods such as Euler-Forward difference in time which uses information from the neighboring cells at present time level $n$ ($U_{i-l}^{n},\ldots,U_{i}^{n},\ldots,U_{i+l}^{n}$) to evolve the solution at time level $n+1$ ($U_{i}^{n+1}$), our architecture uses the entire grid input sequence and encodes it to a context vector and pass it through decoder long short-term memory cell to predict solution at multiple time levels in future ($U^t,\ldots,U^{t+k}$). This facilitates to capture of the propagation of disturbance over long time horizons. The AB-CRAN architecture resembles the time marching capabilities of a multi-point ODE integrator and employs multiple past points to predict the future evolution. Instead of using the fixed weighting coefficient for the past time steps, the AB-CRAN architecture dynamically learns the weighting coefficient through the data that provides an effective capability to predict the solution for a longer time horizon.

\section{Conclusions}
\label{conclusion}
In this work, we have presented a novel attention-based sequence-to-sequence convolutional recurrent autoencoder network for learning wave propagation. 
The challenges of reducing the dimensionality of data coming from hyperbolic PDEs were discussed and the idea of incorporating knowledge within the network architecture has been demonstrated. The proposed AB-CRAN networks serve as an end-to-end nonlinear model reduction tool for wave propagation and convection-dominated flow predictions.
The denoising-based convolutional autoencoder together with the attention-based sequence-to-sequence evolver has been employed as a generalized nonlinear manifold learning for time marching. In relation to the predictive capability, we have demonstrated a remarkable increase in the time-horizon of AB-CRAN in contrast to the standard recurrent neural network with long short-term memory cells on wave propagation problems. Three test problems of increasing complexity namely the 1D linear convection, the 1D nonlinear viscous Burger and the 2D shallow water wave problem were considered to demonstrate the various aspects of the proposed attention-based sequence-to-sequence evolver and the denoising-based convolutional autoencoder. 
We have first assessed the effectiveness of our AB-CRAN algorithm for the linear convection equation. The generality of the present algorithm for the nonlinear phenomenon was successfully demonstrated via the nonlinear viscous Burgers equation. The scalability of our AB-CRAN framework has been successfully shown by solving the 2D Saint-Venant shallow water wave problem.
On both 1D and 2D datasets of hyperbolic PDEs, our novel AB-CRAN with sequence-to-sequence learning accurately captures the wave amplitude and efficiently learns the wave propagation in time. The proposed AB-CRAN framework is general and has the potential to be used for predicting large-scale 3D convection-dominated problems of practical importance.

\section*{Acknowledgements}
The authors would like to acknowledge the funding support from the University of British
Columbia (UBC) and the Natural Sciences and Engineering Research Council of Canada (NSERC).
This research was supported in part through computational resources and services provided by Advanced Research Computing (ARC) at the University of British Columbia and Compute Canada.
\section*{Nomenclature}


\noindent\begin{longtable}{@{}l @{\quad=\quad} l@{}}
AB-CRAN & Attention-based Convolutional Recurrent Autoencoder Net\\
ADAM  & Adaptive Moment Estimation \\
$C$ & Wave Speed \\
$c^{t}, h^{t}$ & States of LSTM Cell t\\
CNN &    Convolutional Neural Network \\
CRAN & Convolutional Recurrent Autoencoder Network \\
FOM   & Full-Order Model \\
K & Convolution Kernel \\
$\mathcal{L}$ & Loss Function \\
$L_\infty$ & Maximum Error \\
LSTM & Long Short-term Memory \\
MSE & Mean Squared Error \\
N & Full-order Spatial Dimensions \\
$N_b$ & Mini-batch Size \\
$\mathcal{N}$ & Normal Distribution \\
$\mathcal{N}_{e}(x)$ & Neighbourhood of $x$ \\
$N_T$ & Total Temporal Length \\
$N_t$ & Input Sequence Length \\
$N_{\mu}$ & Number of PDE Parameters \\
P & Pooling Operator \\
PDE & Partial Differential Equation \\
POD  & Proper Orthogonal Decomposition \\
r & Reduced Dimension \\
$Re$ & Reynolds Number \\
RNN  & Recurrent Neural Network \\
ROM  &  Reduced-order Model \\
$SD$ & Standard Deviation\\
s & Stride Length \\
$\mathcal{U}$ & Full-order Spatial-temporal Data \\
$\mathbf{U}_N$ & Full-order Solution \\
$\mathbf{U}_r$ & Reduced-order Solution \\
X & Model Input\\
Y & Ground Truth\\
$\sigma$ & Non-linear Activation \\
$\Psi_E$ & Encoder Network \\
$\Psi_D$ & Decoder Network \\
$\Phi$ & Temporal Evolver Network \\
$\theta$ & Trainable Parameters\\
$e_{i,t}$ & Soft Alignment Score \\
$\beta_{i,t}$ & Attention Score \\
$h_a$ & Context Vector \\
$\alpha$ & Hyper-parameter Combining Two Losses\\ 
$\mu$ & Parameters of PDE \\
$< >$ & Temporal-average \\
$(\hat{.})$ & Approximation/Reconstruction of Full-order Solution \\
$(\tilde{.})$ & Data with White Noise \\
$\overline{(.)}$ & Min-max Normalized Data \\
$(.^{'})$ & Time-evolved Approximation/Reconstruction \\

\end{longtable}
\section*{Data Availability}
The data that support the findings of this study are available from the corresponding author upon reasonable request.
\section*{Conflict of interest}
The authors declare that they have no conflict of interest.

\section*{References}
\bibliographystyle{plain}
\bibliography{mybibfile}

\end{document}